\documentclass[pra,twocolumn,superscriptaddress,floatfix]{revtex4}
\usepackage{hyperref}
\usepackage{graphicx} 
\usepackage{amsmath}
\usepackage[dvips]{epsfig}
\newcommand{\beq}{\begin{equation}}
\newcommand{\eeq}{\end{equation}} 
\newcommand{\beqa}{\begin{eqnarray}}
\newcommand{\eeqa}{\end{eqnarray}}
\newcommand{\ba}{\begin{array}}
\newcommand{\ea}{\end{array}}

\usepackage{color}

\begin{document}
\title{Vortex properties in the extended supersolid phase of dipolar Bose-Einstein condensates}

\author{Francesco Ancilotto}
\affiliation{Dipartimento di Fisica e Astronomia ``Galileo Galilei''
and CNISM, Universit\`a di Padova, via Marzolo 8, 35122 Padova, Italy}
\affiliation{CNR-IOM, via Bonomea, 265 - 34136 Trieste, Italy }

\author{Manuel Barranco}

\affiliation{Departament FQA, Facultat de F\'{\i}sica,
Universitat de Barcelona.  Av. Diagonal, 645. 08028 Barcelona, Spain}
\affiliation{Institute of Nanoscience and Nanotechnology (IN2UB),
Universitat de Barcelona, Barcelona, Spain.}

\author{Mart\'{\i} Pi}
\affiliation{Departament FQA, Facultat de F\'{\i}sica,
Universitat de Barcelona. Av. Diagonal, 645. 08028 Barcelona, Spain}
\affiliation{Institute of Nanoscience and Nanotechnology (IN2UB),
Universitat de Barcelona, Barcelona, Spain.}

\author{Luciano Reatto}
\affiliation{Dipartimento di Fisica, Universit\`a degli Studi di Milano, 
via Celoria 16, 20133 Milano, Italy  }

\begin{abstract} 

We study the properties of singly-quantized linear vortices 
in the supersolid phase of a dipolar Bose-Einstein 
condensate at zero temperature modeling $^{164}$Dy
atoms. The system is extended in the $x-y$ 
plane and confined by a harmonic trap in the  
the polarization direction $z$.
Our study is based on a generalized Gross-Pitaevskii equation.
We characterize the ground state of the system in terms of spatial 
order and superfluid fraction and compare the properties of 
a single vortex and of a vortex dipole in the superfluid 
phase (SFP) and in the supersolid phase (SSP). 
At variance with a vortex in the SFP, which is 
free to move in the superfluid, a vortex in the SSP is
localized at the interstitial sites and does not move freely. 
We have computed the energy barrier for motion from an equilibrium site to another.
The fact that the vortex is submitted to a periodic potential 
has a dramatic effect on the dynamics  
of a vortex dipole made of  two counter rotating parallel vortices; instead of 
rigidly translating as in the SFP,     
the vortex and anti-vortex approach 
each other by a series of jumps from one site to another
until they annihilate in a very short time and their 
energy is transferred to bulk excitations. 

\end{abstract} 
\date{\today}


\maketitle

\section{Introduction}

Vortices are characteristic quantized topological 
excitations of a superfluid. In the case of a Bose 
superfluid a vortex represents a line singularity 
in the phase of the condensate wave function, the 
phase having an increment of $2\pi$ as one turns around 
this line. Vortices in superfluid $^4$He \cite{donnelly,barenghi} 
and in cold bosonic 
atoms \cite{fetter,Pit16} have been extensively studied over the years both 
as an example of the ubiquitous vorticity 
phenomena in many-body systems as well as a 
signature of the presence of phase coherence in the system. 
Vortices are expected to be also present in a supersolid. 

The supersolid phase (SSP) of matter has attracted considerable interest 
because the two symmetries that are spontaneously broken at the 
same time seem incompatible at first sight \cite{bal10}. On the one hand the 
translational symmetry is broken so that particles localize 
with crystalline order. On the other hand gauge symmetry is broken and this 
leads to a condensate and to the appearance of superfluid properties. 

Supersolidity was 
proposed long ago for solid $^4$He \cite{and69,che70,leg70}. However, experiments have shown 
that low temperature ($T$) solid $^4$He, in spite of displaying a number of anomalous properties, does 
not conform to the bulk supersolid paradigm \cite{cha13,bon12}. Cold bosons have turned out 
to be a more fruitful platform to address supersolidity.  As discussed in the following, 
a number of properties expected for a supersolid have recently been verified for systems made of  
bosonic atoms with a permanent magnetic moment (dipolar bosons).

One should expect differences between the appearance and dynamics of a vortex in a superfluid 
and in a supersolid due to its completely different environment.  
In a superfluid, density can be non-uniform, e.g. close to 
an interface or in the whole volume as for cold atoms in a trap. These are large scale
inhomogeneities compared to the vortex 
core radius, and the vortex is free to move inside the superfluid which it sees as locally uniform. 
This behavior is similar to that of vortices in 
classical hydrodynamics: vortices are free to move inside the fluid
and each vortex moves with the local velocity due to the presence of other vortices
\cite{lamb}. 
The density in a supersolid has a strong spatial modulation 
on the microscopic scale. Since, in order to minimize its kinetic energy, a vortex is a low density 
``seeker'', it is no longer free to move as in the superfluid phase (SFP). 
Rather, it behaves as if it were moving in a
periodic potential with a spatial scale comparable to its core size.
In addition, this density inhomogeneity is not simply due to 
an external potential but the interparticle 
interactions play a key role in establishing it.
This is quite a new regime and the main purpose of 
the present paper is to uncover and discuss some of its properties.

Experiments, in agreement with the predictions of theory, 
show that dipolar bosons can be found in different 
phases. 
When the dipole interaction strength is small, 
the system is in the superfluid phase. 
As the interaction strength increases, the dipolar boson 
dispersion relation 
(excitation energy as a function 
of the momentum of the excitation, ${\cal E}(k)$) displays
a roton minimum  whose gap amplitude depends
on the relative strengths of short-range and dipolar interactions \cite{lewe,odell}.
Such dispersion relation has some similarities with that 
of superfluid $^4$He \cite{donnelly,Pit16}.
When the ratio of the dipolar to the short-range interaction
increases beyond a critical value, 
a spontaneous density modulation occurs
which is driven by the softening of the 
roton mode at finite momentum $k_R$,
and the resulting system has supersolid character
\cite{wenzel,blackie_baillie3,Chomaz2018,lewe,kora,ancilotto}.
The density modulation, with wavelength $\sim 2\pi/k_R$, 
results in an ordered array of 
``droplets'',  made of many atoms each, elongated in the direction of the polarization axis. 
Global phase coherence is maintained between adjacent droplets
due to a low-density superfluid background that
allows atom tunneling from one superfluid droplet to another.
As the ratio between interactions is further increased, 
the superfluid background almost disappears
so that phase coherence is no longer present between droplets,
and the phase becomes a ``normal solid phase'' (NSP in the following),
even if each droplet can be superfluid.

Further increase of that ratio
eventually results
in a transition to a totally different regime, characterized by
the formation of
{\it self-bound} droplets \cite{Kad16,Fer16,schmitt,santos_wachtler,blackie_baillie1,ferlaino},
with order-of-magnitude higher densities, the binding
arising from the interplay between the
two-body dipolar interactions and the effect of quantum
fluctuations \cite{lima_pelster,santos_wachtler}.

In dipolar bosons supersolid behavior occurs in the intermediate 
regime between superfluid and self-bound droplets regimes.
We remark that the term droplet will 
be used here to indicate the individual clusters making
up the ordered structure of a modulated phase in such intermediate regime,
and will not refer to the self-bound droplet regime.

A number of theoretical studies predicted supersolid behavior in dipolar 
Bose-Einstein condensates (BEC) in different geometries
\cite{bombin,cinti1,wenzel},
dipolar gases confined
in a quasi-2D pancake shaped trap \cite{blackie_baillie3,blackie_baillie1},
or in a tube \cite{ancilotto}.
The order of the superfluid-supersolid transition has been studied 
in Ref. \cite{pohl}.

Experiments provide evidence of SSP
in dipolar bosons in a number of ways.
Strong global phase coherence was found in
the SSP realization of Ref. \cite{chomx} 
as opposed to the lack of it in the 
isolated droplet phase where no superfluid flow
is present between adjacent droplets in the NSP.
Similarly, robust phase coherence across a 
linear array of quantum droplets
was observed in Ref. \cite{bott19}.
Stable stripe modulations have also been observed
in dipolar quantum gases \cite{wenzel,tanzi}.
A partial phase coherence is suggested in Ref. \cite{tanzi},
thus indicating possible SSP behavior.
The characteristic symmetry breaking of a SSP was observed 
through the appearance of compressional oscillation 
modes in a harmonically trapped dipolar condensate \cite{tanzi_new}.
In another work \cite{tanzi19}, the reduction of 
the moment of inertia under slow rotation --previously predicted for
a dipolar SSP \cite{stringa19}-- has been measured.
The response of the dipolar SSP has been 
studied experimentally in Ref. \cite{ferla19}, where 
the out-of-equilibrium superfluid flow 
across the whole system was revealed by a rapid re-establishment of 
global phase coherence after a phase-shattering excitation was applied. Instead, no such 
re-phasing was observed in the NSP, where tunneling between adjacent droplets is suppressed.

Most of the recent evidence of supersolid 
behavior in dipolar gases is based on 
the identification of two main features of a SSP,
i.e. (i) a non-zero non-classical translational/rotational inertia \cite{sep_joss_rica}
and (ii) the appearance  of the
Nambu-Goldstone gapless mode \cite{pfau} corresponding to phase fluctuations
--besides the phonon mode associated to density fluctuations which results from
the translational discrete symmetry of the system.
But no evidence of another hallmark of 
superfluidity has been gathered so far, 
namely the presence of quantized vortices \cite{stringa19}.
Experimental realization and detection of quantized vortices would provide 
a direct evidence of global coherence 
in the SSP of a dipolar system.

The structure of vortex arrays in a trapped supersolid system
rotating in the plane orthogonal to the polarization axis
has been theoretically studied in Ref. \cite{stringa20}, where
vortex nucleation was found to be 
triggered, as in standard condensates,
by the softening of the surface quadrupole mode.
For larger values of the angular velocity of the rotating trap
an array of regularly arranged vortices appears,
which coexists with the triangular geometry of the supersolid
lattice and persists during the free expansion of the
atomic cloud.
The vortex cores are localized in 
the interstitial regions between droplets where the 
local particle density can be very small.
As a consequence, the critical angular velocity for the nucleation
of an energetically stable vortex in the SSP has been found
to decrease with respect to that in the SFP \cite{stringa20}.
It was shown in Ref. \cite{arxiv}  that a vortex-hosting trapped SSP 
could be obtained by starting 
from the vortex-hosting superfluid phase  and  
tuning the interaction strength to that of 
the SSP. There it was also shown that evidence 
for the presence of vortices in the SSP could be 
obtained from interference features of the expanding cloud.

Previous theoretical studies of vortices in a SSP 
have been performed for dipolar bosons in full confinement 
in a rotating trap, i.e. in a harmonic trap along the three 
spatial directions. In order to achieve an understanding 
of the intrinsic properties of a vortex in the SSP, 
in the present work we study isolated vortices in 
the SSP of an extended dipolar BEC, i.e. free 
from the radial harmonic trap in the plane perpendicular 
to the polarization direction but subject to a harmonic 
confinement along it. The use of periodic boundary conditions 
in the plane dictates that the crystalline structure of 
the SSP is fixed in space and vortices are imprinted by 
suitable phase factors. We find that the isolated vortices 
are stable excitations of the system. 

We compare the properties of a vortex in the SFP of 
dipolar bosons with those in the SSP and study 
how the vortex energy changes depending on the 
location of the vortex core in the droplet network. 
We find that a vortex dipole, i.e. two vortices of opposite
chirality, has a completely different
behavior in the SSP than in the SFP.
A vortex dipole in the SFP uniformly translates
keeping a fixed inter-vortex distance unless the two 
vortices are very close at distances of order of the 
healing length \cite{donnelly,roberts}. 
At variance, 
as a consequence of the spatial modulation 
of the condensate, we find that in the SSP the dipole does not translate and 
it is no longer a 
stable excitation of the system; rather, the two vortices 
of opposite chirality feel an attractive interaction 
so that they approach each other until they
annihilate each other and their energy is 
transferred into bulk excitations.

This paper is organized as follows.
In Sect. II we present the theoretical model and outline the 
numerical methods used in this study. 
The ground state of dipolar 
bosons in the SFP and in the SSP is studied in Sect. III. We present the 
results for a single vortex in Sect. IV, and 
discuss the results for a vortex dipole in Sect. V.
Finally, Sect. VI contains a summary 
of our study.

\section{Method}

A standard approximation for dipolar bosonic atoms with
mass $m$ and magnetic moment $\mu$  at $T=0$ is represented by a macroscopic wave function
$\phi({\bf r})$ that obeys the extended Gross-Pitaevskii equation (eGPE) \cite{santos_wachtler}:

\begin{align}
 H\phi({\bf r})\equiv
& \left\{-\frac{\hbar^2}{2m}\nabla^2+V_t({\bf r})+g|\phi({\bf r})|^2+
\gamma(\epsilon_{dd})|\phi({\bf r})|^3+ \right.
\nonumber
\\
  & \left. \int d{\bf r'}|\phi({\bf r'})|^2V_{dd}({\bf r}-{\bf r'})
   \right\}\phi({\bf r})=\varepsilon  \phi({\bf r})
 \label{acca}
\end{align}
Here $g=4\pi\hbar^2a_s/m$, $a_s$ being the s-wave scattering length,
$V_{dd}({\bf r})=\frac{\mu_0\mu^2}{4\pi}\frac{1-3\cos^2\theta}{r^3}$
is the dipole-dipole interaction between two identical magnetic dipoles
aligned along the z axis
($\theta$ being the
angle between the vector ${\bf r}$ and the polarization direction $z$), and
$\mu_0$ is the permeability of the vacuum.
$V_t({\bf r})$ is the trapping potential.
The number density of the system is $\rho ({\bf r})=|\phi ({\bf r})|^2$.
The $\gamma(\epsilon_{dd})$ term
is the beyond-mean-field (Lee-Huang-Yang, LHY) correction \cite{lima_pelster},
where $\gamma(\epsilon_{dd})=
\frac{32}{3\sqrt{\pi}}ga_s^{3/2}F(\epsilon_{dd})$,
$\epsilon_{dd}=\frac{\mu_0\mu^2}{3g}$
being the ratio between the strengths of the dipole-dipole and
contact interactions, and
$F(\epsilon_{dd})=\frac{1}{2}\int_0^{\pi}d\theta
\sin\theta[1+\epsilon_{dd}(3\cos^2\theta-1)]^{5/2}$.
The chemical potential $\varepsilon$ is determined
by the normalization condition 
$\int |\phi ({\bf r})|^2 d{\bf r}=N$, $N$ being the total number of dipoles.

We solve the above equation by propagating  it
in imaginary time, if stationary states are sought, or by
propagating in real-time its time-dependent
counterpart $i\hbar \partial \phi/\partial t=H\phi$ to
simulate the dynamics of the system,
starting from a suitable initial wave function specified in the
following Sections.

To compute the spatial derivatives
appearing in Eq. (\ref{acca})
we used an accurate 13-point finite-difference formula \cite{Anc17}.
The convolution integral in the potential energy term
of Eq. (\ref{acca}) is efficiently evaluated
in reciprocal space by using Fast Fourier transforms,
recalling that the Fourier transform of the dipolar interaction is 
${\tilde V}_{\bf k}=(\mu _0 \mu/3)(3\,cos^2\alpha -1)$,
where $\alpha$ is the angle between ${\bf k}$ and
the $z$-axis \cite{lahaye}.
The spatial mesh spacing and time step are chosen such that 
during the time evolution excellent conservation
of the total energy of the system is guaranteed.
We verified that the  
dimension of the simulation cell along the $z$-direction,
where the harmonic confinement is active,
is wide enough to make negligible  
the effect, on the energy values  
and density profiles, of the 
dipole-dipole interaction between periodically repeated images.

The investigated system is made by $N \sim 1.3-1.4\times 10^6$ $^{164}$Dy atoms
(the chosen $N$ value slightly changes depending upon 
the value of $a_s$, as discussed in the following)
subject to a harmonic trapping potential along the $z$ axis,
with frequency $\omega_z=150 \times 2\pi \,$ Hz.
The harmonic length associated to this trapping potential is 
$a_{ho}=\sqrt{\hbar^2/m\omega_z}=1.2112\times 10^4\,$a$_0=0.64\,\mu m$.
The dipolar gas is thus extended in the $x-y$ plane
(with periodic boundary conditions), and flattened in 
the direction of the dipole polarization.
Typical transverse sizes 
are $L_x,L_y\sim 30\,\mu m$.

A linear, singly quantized vortex excitation in
the extended dipolar system with axis in 
the $z$-direction and core in the position $(x_v,y_v)$,
can be generated by the ``imprinting procedure'' \cite{Anc17}, 
i.e. we compute the lowest energy state
obtained by  starting the imaginary time evolution from
the wave function
\begin{equation}
\phi_0(\mathbf{r})=\rho_0^{1/2}(\mathbf{r})\, \left[ {(x-x_v)+\imath (y-y_v) 
\over \sqrt{(x-x_v)^2+(y-y_v)^2}}  \right]
\label{imprint}
\end{equation}
where $\rho_0(\mathbf{r})$ is the
ground-state density of the vortex-free dipolar system. 
During the imaginary time evolution,
the vortex position and vortex core structure 
change to provide at convergence the lowest energy
configuration.
 
The flow field of a linear vortex has a long-range character, 
$\sim 1/r$, where $r$ is the distance from the vortex axis. 
We have imposed antiperiodic boundary conditions \cite{Pi07} in order 
that the condition of no flow across the boundary of 
the computational cell  is satisfied. This is equivalent 
to sum over the phases of an infinite array of vortex-antivortex, 
i.e. a vortex of opposite chirality is present in each 
nearest neighbor cell of the computation cell \cite{sadd}. 
The sign of the circulation can be changed by changing the sign of the complex 
$\imath$ in Eq. (\ref{imprint}); this equation can be 
easily generalized to accomodate a vortex array made of an arbitrary number
of vortices and/or antivortices \cite{Anc18}. 
A generalization to the case of a pair of vortices with opposite circulation
(vortex dipole) will be used in Sec. V.

\section{Ground-state in dipolar bosons}

We have taken as a case of study a dipolar bosons system made of $^{164}$Dy atoms.
The relative strength of the dipolar interaction 
over the short-range one is defined by the dimensionless parameter
$\epsilon _{dd}=a_{dd}/a_s$, where $a_{dd}=132$\,a$_0$ is the 
dipolar length for $^{164}$Dy atoms \cite{ferla19}.
The behavior of dipolar bosons depends crucially on the
value of the $\epsilon_{dd}$ ratio, whose value 
can be experimentally varied by changing $a_s$.

When $\epsilon_{dd}$ is
below a threshold value $\epsilon ^c_{dd} $ the
system at $T=0$ K is in a standard superfluid state,
in our case it is uniform in the $x-y$ plane and non-uniform
in the $z$-direction due to the trap in that direction.
When $\epsilon _{dd}$ is above the threshold the system spontaneously 
develops a periodic density modulation in the $x-y$ plane, entering
the SSP.

\begin{figure}[!]
\centerline{\includegraphics[width=1.0\linewidth,clip]{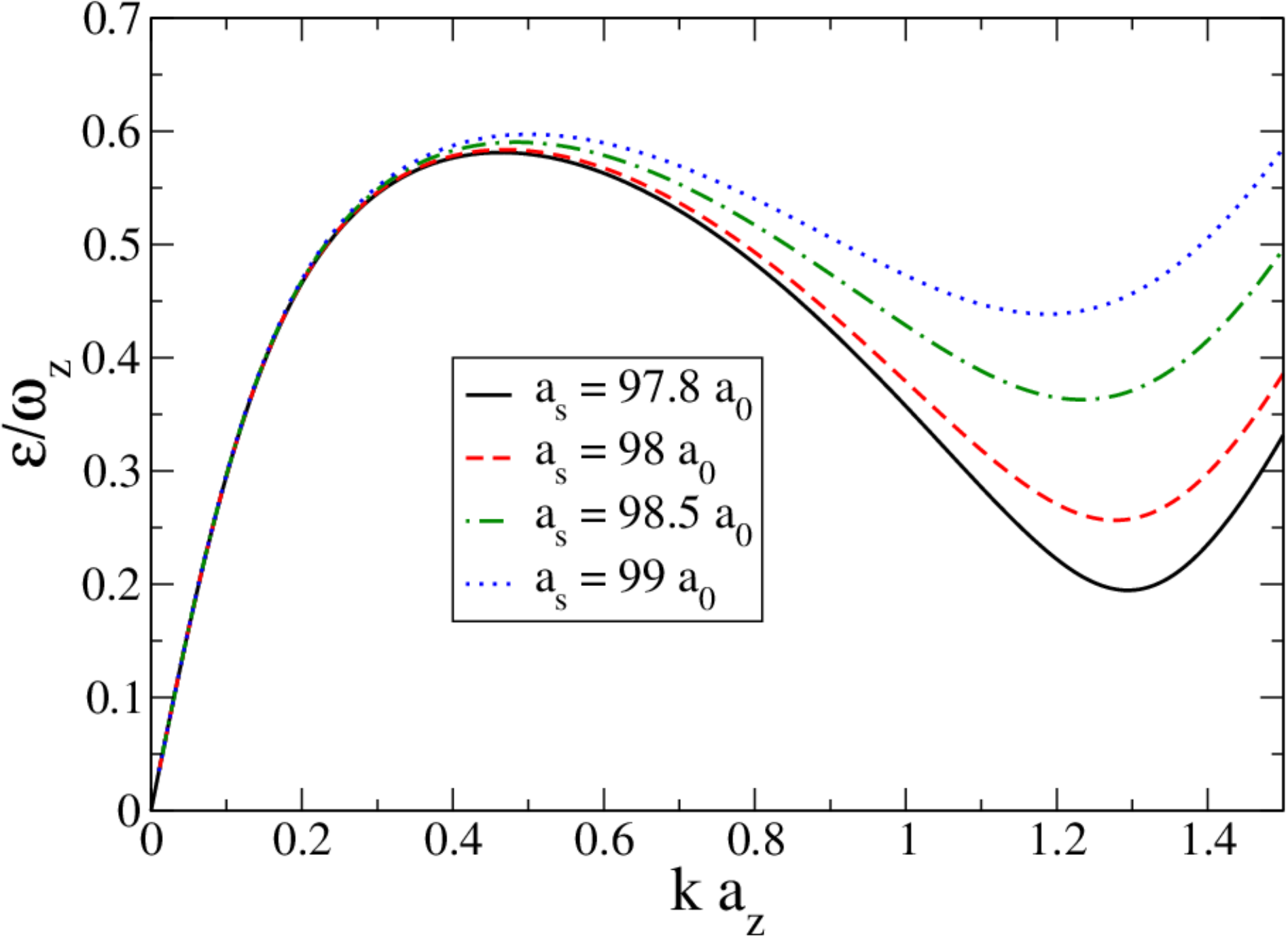}}
\caption{Dispersion relation ${\cal E}(k)$ of the extended SFP
as a function of the wave vector $k$ in the $x-y$ plane for different 
values of $a_s$. ${\cal E}(k)$ is in units of the
harmonic energy $\omega_z$ and $k$ is in units of the 
inverse of the harmonic length $a_{ho}=\sqrt{\hbar/(m\omega_z)}$.
}
\label{fig1-SM}
\end{figure}

As $\epsilon_{dd}$ is increased toward the threshold,
one can get a signature of the SFP to SSP transition from 
the excitation spectrum of the superfluid, ${\cal E}(k)$. 
We first obtain the ground state SFP wave function 
by propagating in imaginary 
time a starting wave function which is just a constant in the $x, y$ 
variables times the ground state wave function for free
particles in the $V_t(z)$ trap. 
Next, using the method described in Ref. \cite{ancilotto}, we solve
the Bogoliubov-de Gennes  
equations for superfluid states corresponding to different $\epsilon _{dd}$ values.
For small $\epsilon _{dd}$ the dispersion relation ${\cal E}(k)$ is structureless, 
starting as a linear function of $k$ corresponding to phonon excitations which evolves toward 
a quadratic dispersion at large $k$ corresponding to  free particles. 
Increasing $\epsilon _{dd}$, the excitation spectrum 
changes shape, and close to $\epsilon^c_{dd}$ ${\cal E}(k)$
develops a phonon-maxon-roton structure as shown in Fig. \ref{fig1-SM}.
This is due to the interplay between confinement in the 
$z$-direction and the interatomic interaction consisting 
of the short range interaction plus the long-range 
interaction of the dipoles oriented in the direction of the confinement \cite{lewe,odell}.

The four curves in Fig. \ref{fig1-SM}
show the excitation spectrum of the SFP for $a_s=97.8,\, 98,\, 98.5$, and 99\,a$_0$,
which correspond to a narrow range of $\epsilon _{dd}$ values between
$\epsilon _{dd}= 1.35$ (corresponding to $a_s=97.8$\,a$_0$) and $\epsilon _{dd}= 1.33$
(corresponding to $a_s=99$\,a$_0$).
The vanishing of the roton gap in the present case occurs just below $a_s=97.8$\,a$_0$,
corresponding to the value $\epsilon^c_{dd}\sim 1.35$.
Notice that this value depends on the strength of confinement 
in the $z$-direction and on the mass of the bosons.
Beyond the threshold the system becomes non-uniform in the $x$ and $y$-directions.

We obtain the ground state wave function in the SSP
by evolving in imaginary time a starting wave function that is 
a superposition of gaussian profiles 
in the $x-y$ plane, located at the sites of a triangular lattice, modulated
along the $z$-direction by the ground state of free 
particles in the harmonic trap in that direction.
In the following, most results are obtained with $\epsilon_{dd}$ 
in the $1.39<\epsilon _{dd}<1.48$ range, where the system 
shows spontaneous density modulation with supersolid behavior.
The scattering lengths associated to this interval 
are $95\,a_0 > a_s > 89\,a_0$.

\begin{figure}[t]
\centerline{\includegraphics[width=1.1\linewidth,clip]{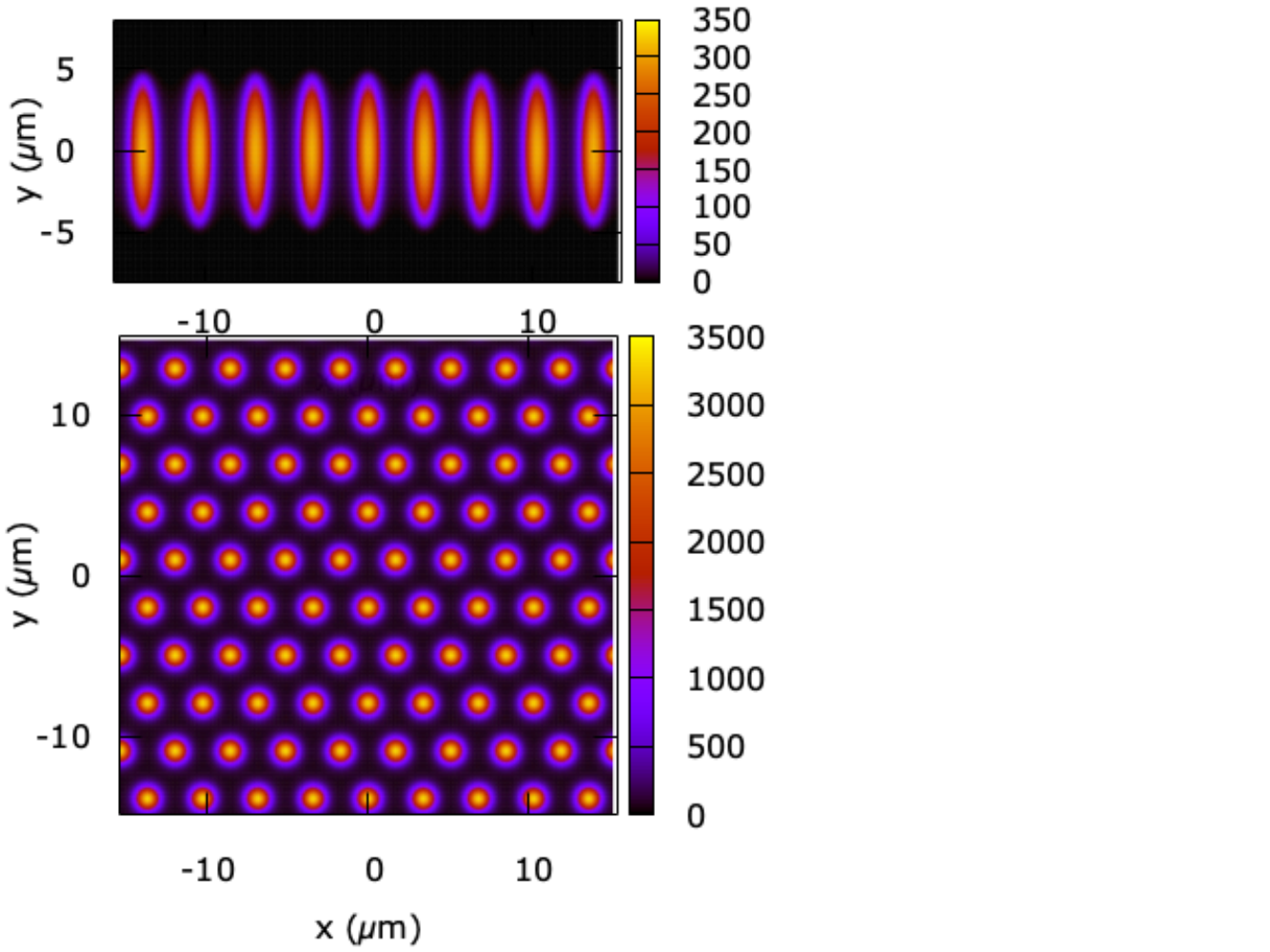}}
\caption{Supersolid phase for the extended system with $a_s=93\,a_0$.
Upper panel:
number density $\rho$ in a $x-z$ plane passing
through a row of droplets, in units of $a_{ho}^{-3}$.
Lower panel: integrated density along the $z$-direction, 
shown in the $x-y$ plane,
in units of $a_{ho}^{-2}$.
The lengths along the axis are in $\mu m$.
}
\label{fig1}
\end{figure}

The modulated system obtained in this way 
consists of a planar ordered 
array of ``droplets'' with triangular symmetry, 
elongated in the direction of the polarization axis, 
each containing $\sim 15{,}000$ atoms.
To good approximation, these cigar-shaped droplets 
are circular in the $x-y$ plane and with an extension 
in the $z$-direction  which is typically 4-5 times larger than their diameter
in the $x-y$ plane, as shown in Fig. \ref{fig1} for $a_s=93\,a_0$.

An almost uniform fluid occupies the 
space between droplets, and its density 
is much smaller than that of the droplets. 
Global phase coherence is maintained between adjacent 
droplets due to this low-density superfuid background
which allows tunneling of atoms from one droplet to another. 
Typical average densities in the $z=0$ plane of symmetry
are $\rho _{av}\sim 3.5\times 10^{-11}\,a_0^{-3}$.

\begin{figure}[t]
\centerline{\includegraphics[width=1.0\linewidth,clip]{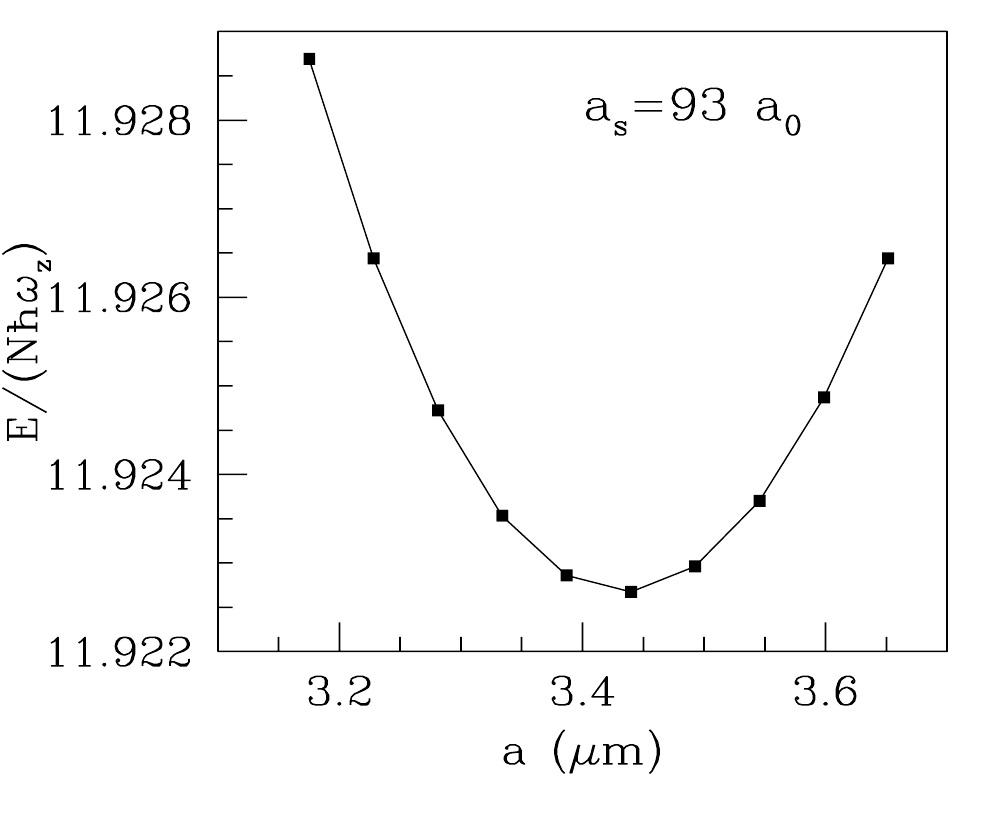}}
\caption{ Energy per atom vs. $a$ for the case $a_s=93\,a_0$.
}
\label{fig2}
\end{figure}

\begin{figure}[t]
\centerline{\includegraphics[width=1.0\linewidth,clip]{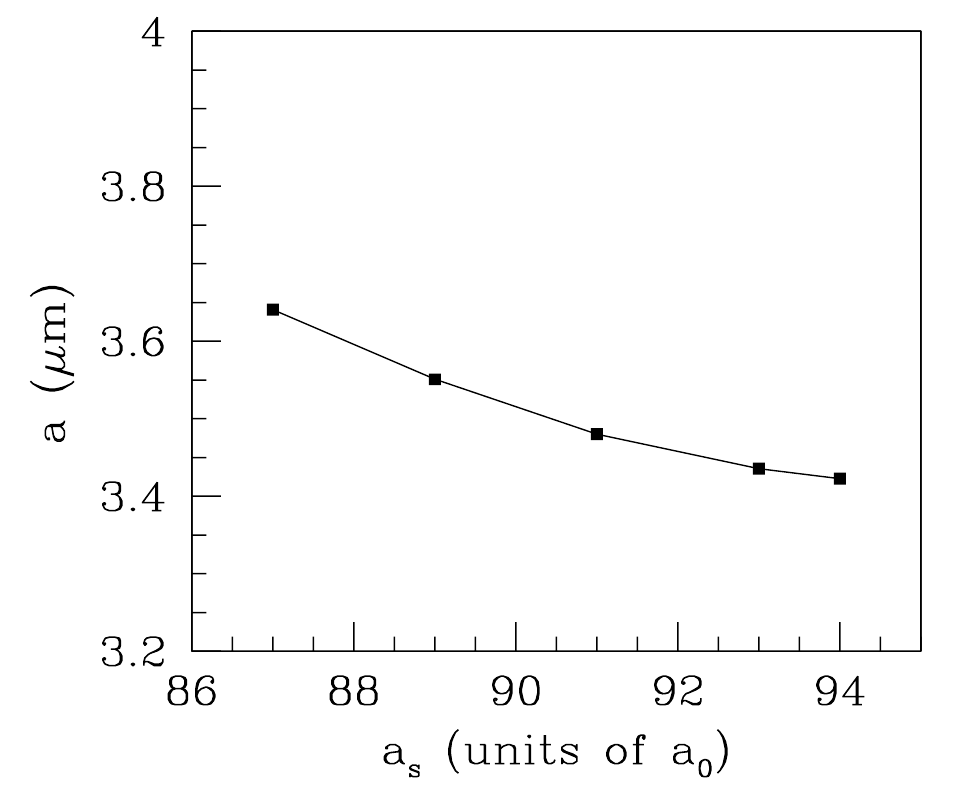}}
\caption{ Equilibrium value of the lattice constant $a$
as a function of the scattering length $a_s$. 
}
\label{fig3}
\end{figure}

The triangular packing of droplets is characterized by a lattice constant
$a$, corresponding to the nearest-neighbor distance
between two adjacent droplets.
The periodic boundary conditions impose a 
constraint on the value of $a$ such that an 
integer number of crystalline cells 
fits in the size of the computation box. 
For a given value of the scattering length $a_s$, 
we preliminary determine the
equilibrium value for $a$ by computing the energy per atom
for different values of $a$ and look for the minimum
energy value.
An example of such calculation is shown in Fig. \ref{fig2},
where the energy per atom is shown --for the $a_s=93\,a_0$ case-- 
as a function of the lattice constant $a$.
As a result, the total number $N$ of atoms 
depends on the value of $a_s$, and $N$ changes by 
about 8\% in the range of scattering length 
considered in our work.
Fig. \ref{fig3} shows the equilibrium values of the lattice constant 
for different choices of $a_s$.

All the modulated structures investigated here
have supersolid character.
In order to prove this we checked for the presence 
of a finite non-classical translational inertia (NCTI)
in the system.
This is done by solving for stationary states the
real-time version of Eq. (\ref{acca}) in the
comoving reference frame with uniform velocity $v_x$, i.e.
\begin{equation}
 i\hbar\frac{\partial}{\partial t}\phi({\bf r})=
\left(H+i\hbar v_x\frac{\partial}{\partial x}\right)\phi({\bf r})
\end{equation}
Following Ref. \cite{sep_joss_rica}, we define the superfluid fraction $f_s$
as the fraction of particles that
remains at rest in the comoving frame:
\beq
 f_s=1-\lim_{v_x\to 0} \frac{\langle P_x \rangle}{N m v_x}
\eeq
where $\langle P_x \rangle =-i\hbar \int d \mathbf{r}\, \phi ^{\ast} \partial \phi /\partial x$
is the expectation value of the momentum and $Nmv_x$ is
the total momentum of the system if all atoms were moving.
We have verified that, in the low-velocity limit, the same value for $f_s$
is obtained by applying instead the boost along the $y$-direction.
We remark that the superfluid fraction computed in this way 
should not be confused with the total superfluid fraction
because each droplet is superfluid but do not 
necessarily contribute to the superfluid response of the overall system. 

\begin{figure}[t]
\centerline{\includegraphics[width=1.0\linewidth,clip]{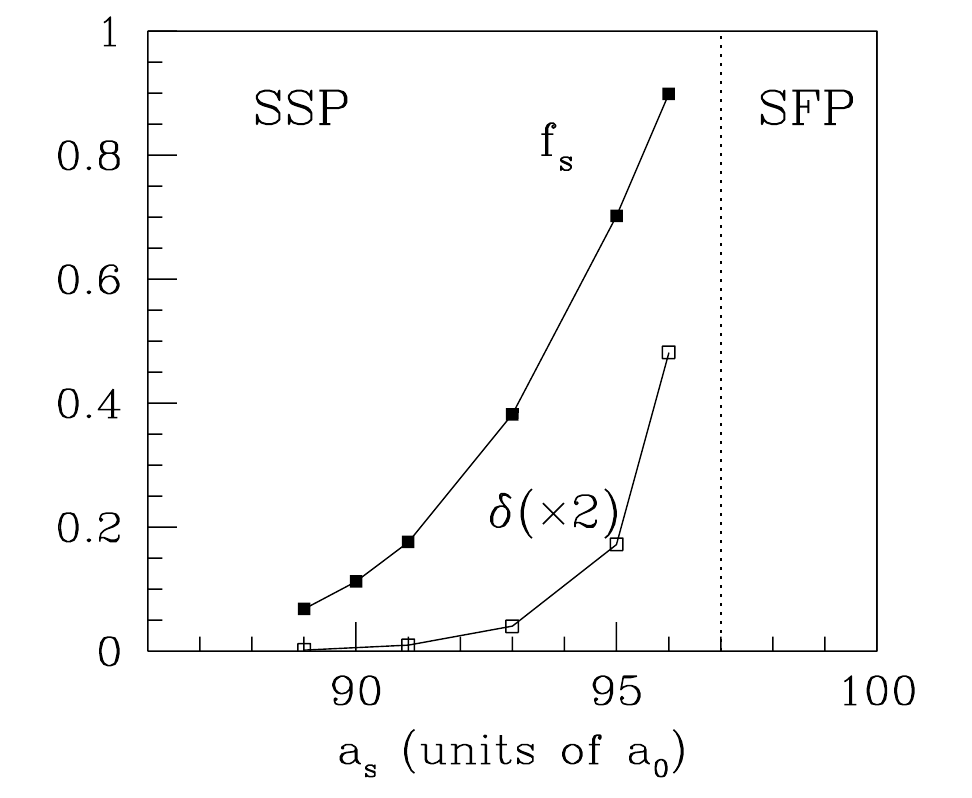}}
\caption{ Upper curve: superfluid fraction $f_s$. Lower curve: 
ratio $\delta = \rho _{min}/\rho _{max}$.
The vertical line shows the location of the SFP $\rightarrow$ SSP transition.
}
\label{fig4}
\end{figure}

The calculated average superfluid fraction is shown 
in Fig. \ref{fig4} as a function of $a_s$.
We also show in the same figure the ratio 
between the density $\rho _{min}$ of the superfluid background
in the interstitial site between three adjacent droplets 
and the maximum density of the clusters in the $z=0$ plane of symmetry, 
$\delta = \rho _{min}/\rho _{max}$.
Not surprisingly, the lower the ratio, the 
smaller the associated superfluid fraction.
We find that the SFP $\rightarrow $ SSP transition 
occurs around the value $a_s=97\,a_0$ (an exact determination 
from the simulation is problematic: close to the transition
residual, very low modulations in the density 
take overly long imaginary times to be reduced). 
The vertical line in Fig. \ref{fig4} shows the estimated 
location of 
the SFP $\rightarrow $ SSP transition. For values of $a_s$
to the right of this line, the lowest energy structure is
a homogeneous (in the $x-y$ plane) SFP with $f_s=1$.

In the SSP regime, when the value of $a_s$ is progressively decreased the  
background density becomes steadily smaller and eventually the 
system is no longer supersolid, being composed 
of isolated droplets with incoherent phases \cite{chomx}.
We cannot follow this transition from a supersolid 
to this ``normal solid'' made of droplets (the NSP in our notation) 
because the theory we use, which is based on a single order parameter 
implying a global phase coherence across the whole system, is not 
fully appropriate when the different droplets become incoherent. 

\section{Single vortex in dipolar bosons}

\begin{figure}[!]
\centerline{\includegraphics[width=0.95\linewidth,clip]{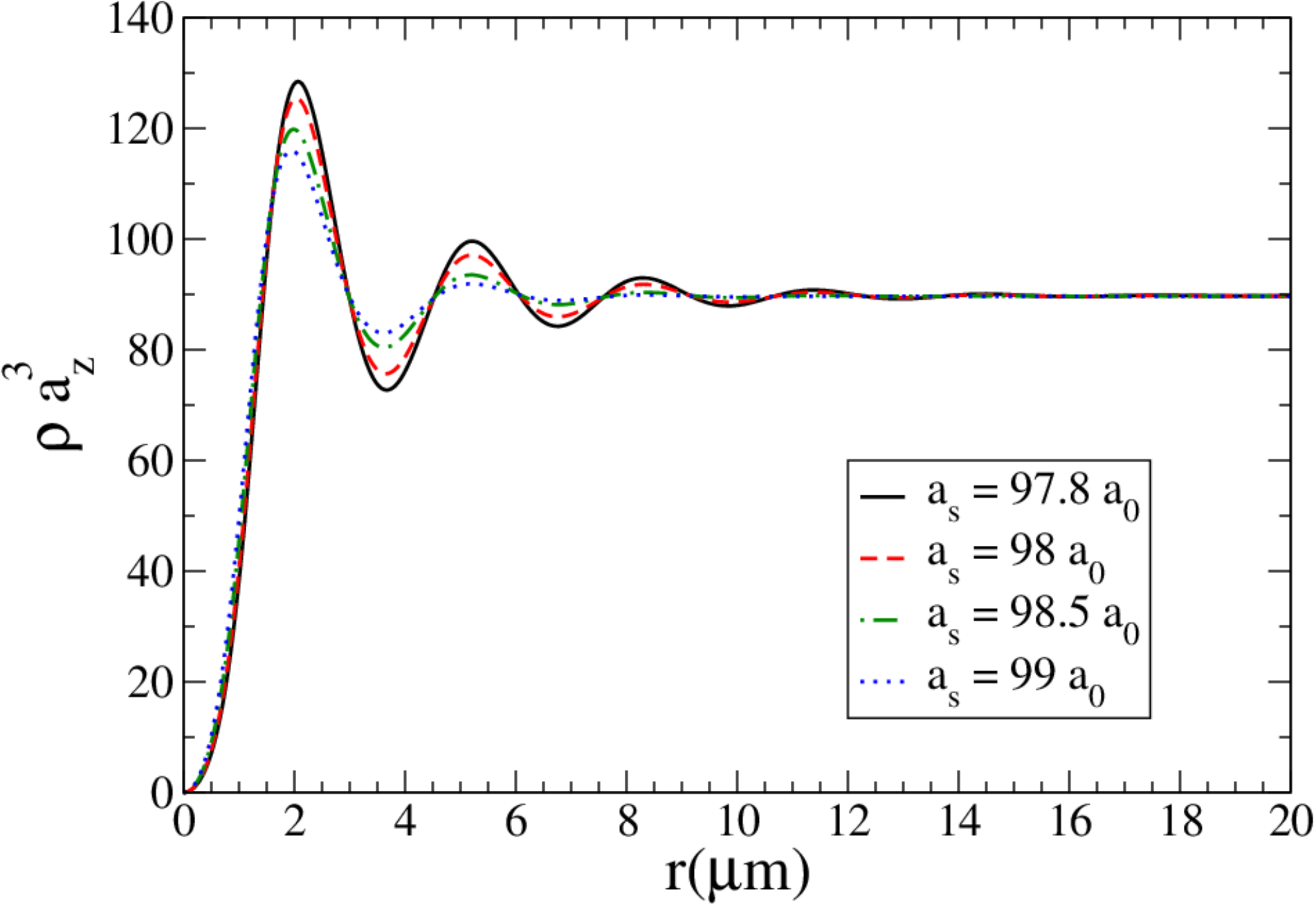}}
\caption{Radial vortex density profiles in the SFP for different $a_s$ values.
}
\label{fig2-SM}
\end{figure}

We have studied singly quantized vortices in our
extended system for different values of the scattering length $a_s$.
In the SFP, the density profile is featureless for large $a_s$ values, 
quadratically vanishing at the position of the vortex core 
and monotonously approaching the bulk density at large distances,
as in usual BEC systems. As $a_s$ is decreased (and thus $\epsilon _{dd}$ increases) 
the density profile develops damped oscillations near the vortex core \cite{Yi06}. 
This is shown in Fig. \ref{fig2-SM}, where we display the equilibrium vortex-hosting density profiles 
in the SFP for the same $a_s$ values as in Fig. \ref{fig1-SM}.
The smaller the $a_s$, the higher the density peaks around the vortex core. 
The oscillations become  more pronounced as the instability limit $\epsilon^c _{dd}\sim 1.35$ is approached. 
The wave vector of the oscillations in the vortex-hosting density profile 
is close to the roton $k_R$.
These oscillations are similar to those predicted in superfluid $^4$He which  
have been described as a cloud of virtual rotons \cite{reatto}. 

\begin{figure}[t]
\centerline{\includegraphics[width=0.8\linewidth,clip]{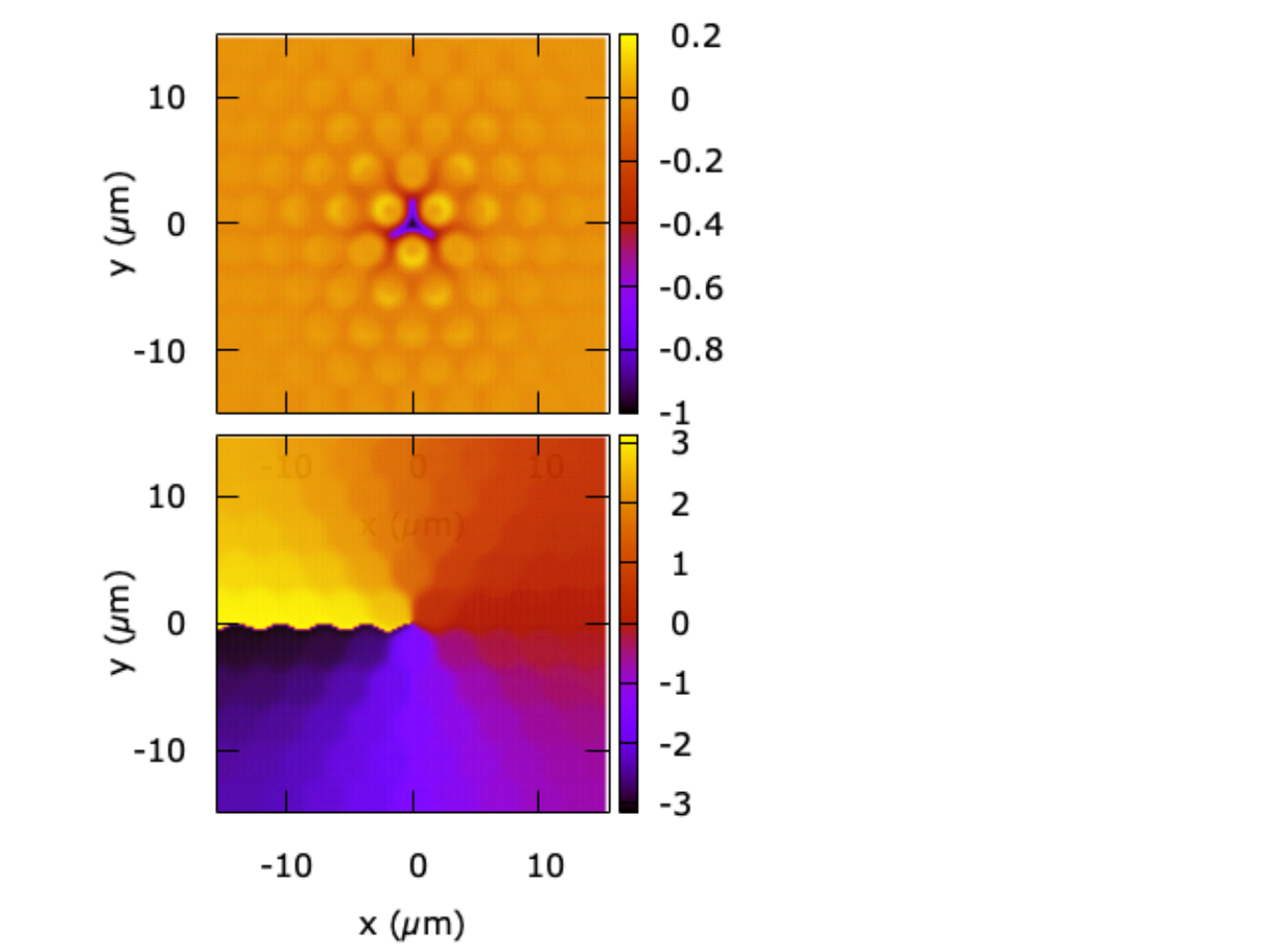}}
\caption{ Upper panel: 
Density difference 
($\rho _{vortex} -\rho_{no vortex})/\rho_{no vortex}$ in the $z=0$ symmetry plane 
for the vortex at the interstitial site when $a_s=93\,a_0$.
Lengths are in $\mu m$.
Lower panel:
Phase (in radians) of the vortex state shown in the upper panel.
The lengths along the axis are in $\mu m$.
}
\label{fig4b}
\end{figure}

As we have discussed previously, 
when $a_s$ becomes smaller than a critical value the system
enters the supersolid phase, as in the structure shown in
Fig. \ref{fig1}.
In this phase, the lowest energy position of the vortex core is 
in the interstitial region between three adjacent droplets, where the 
local particle density $\rho ({\bf r})$ is very small, so that
the energy cost of the additional
depression of $\rho ({\bf r})$ near the vortex core 
(remember that the local density has to vanish at the vortex core) is minimized and  
the kinetic energy of the superfluid flow around the vortex core also decreases
\cite{stringa19}.
To visualize the vortex core structure, the map of the
difference between the densities of the configuration with and
without vortex at the interstitial site for the $a_s=93\,a_0$ case  
is shown in the upper panel of Fig. \ref{fig4b}.
It appears that the core is very deformed 
following the symmetry of the interstitial site \cite{rocc}.
We have found a similar vortex core structure for other values of the scattering length.

\begin{figure}[t]
\centerline{\includegraphics[width=1.0\linewidth,clip]{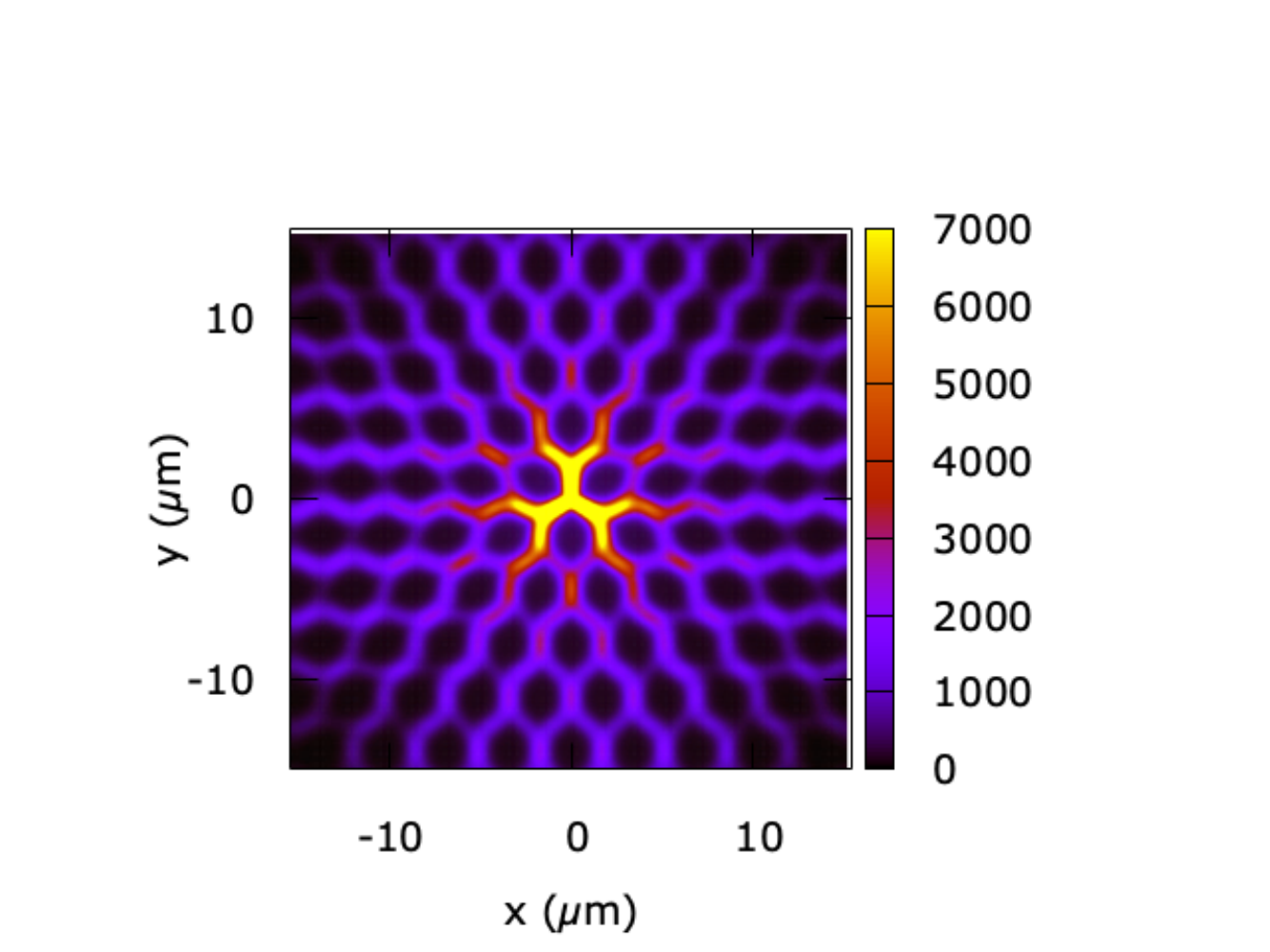}}
\caption{
Modulus of the velocity field (in units of $a_0 \omega_z$) 
on the 
$z=0$  plane of symmetry for $a_s=93\,a_0$.
The lengths along the axis are in $\mu m$.
}
\label{fig11}
\end{figure}

The phase associated to the vortex 
is shown in the lower panel of Fig. \ref{fig4b}.
Notice the hexagonal texture which is a consequence of the
triangular lattice of clusters in the vortex-hosting SSP.
As shown in Fig. \ref{fig11},
this texture becomes prominent in the plot of 
the modulus of the velocity field $|{\bf v}({\bf r})|$, where 
${\bf v}({\bf r})=(\hbar/m)\nabla _R\theta({\bf r})$ with
$\theta $ being the phase of the order parameter $\phi$,
and $\nabla _R\equiv (\partial/\partial x,\partial/\partial y)$.
It appears that $|{\bf v}({\bf r})|$ completely
differs from the $1/r$ behavior in a superfluid, being instead 
strongly anisotropic and modulated, and very small
inside the droplets. We will show below how
this is affecting the kinetic energy of the vortex.

\begin{figure}[t]
\centerline{\includegraphics[width=0.9\linewidth,clip]{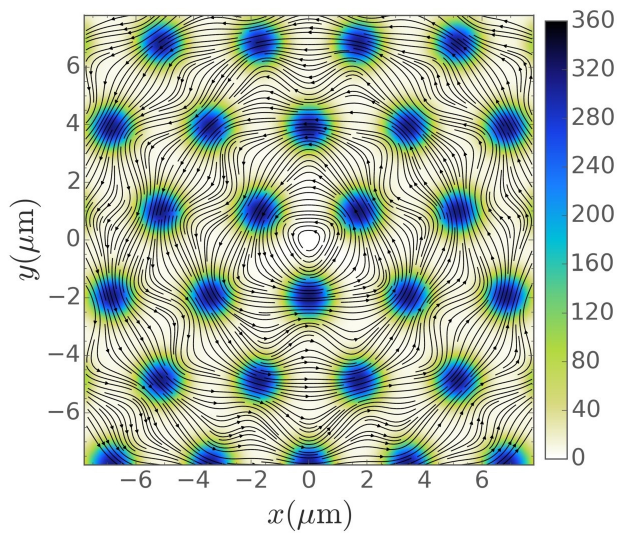}}
\caption{Velocity streamlines for the SFP state for $a_s=93\,a_0$.
The vortex core is in the origin. 
Superimposed to the field lines is the density in the $z=0$ plane, in units of 
$a_{ho}^{-3}$.
}
\label{stream1}
\end{figure}

We show in Fig. \ref{stream1} the velocity streamlines in the $z=0$ symmetry plane
for the case described above, which are remarkably different from the
simple tangential flow of a vortex in a homogeneous superfluid.
We find that the circulation calculated around a closed path encircling the
vortex core is $\oint {\bf v}\cdot {\bf dl}=(2\pi \hbar /m)n$, with $n=1.01\pm 0.01$, and
where the error reflects slightly different 
outcomes depending on the chosen size of the closed path around the vortex core.

\begin{figure}[t]
\centerline{\includegraphics[width=1.0\linewidth,clip]{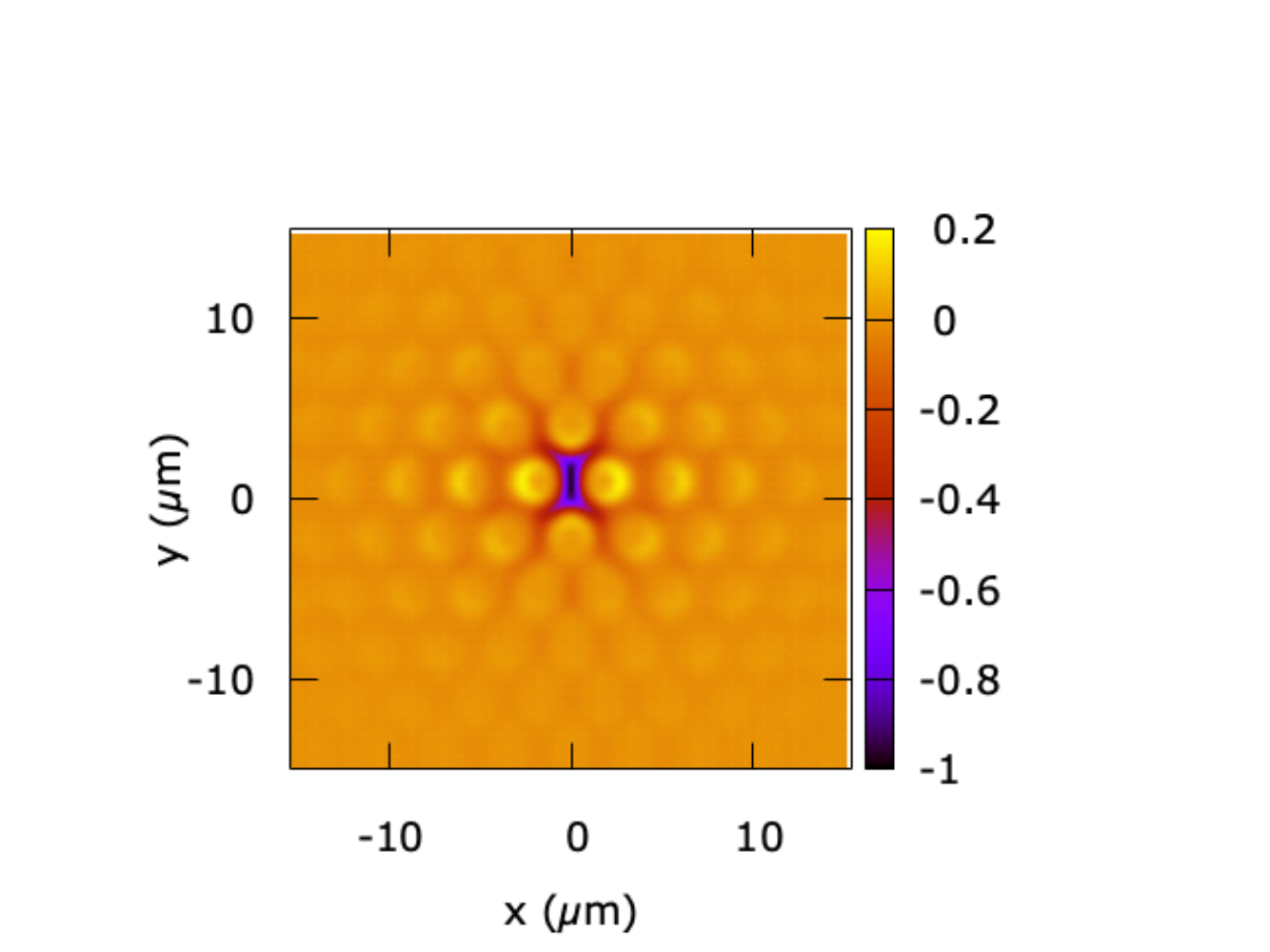}}
\caption{Density difference 
($\rho _{vortex} -\rho_{no vortex})/\rho_{no vortex}$ in the $z=0$ symmetry plane 
for the vortex at the saddle-point site when $a_s=93\,a_0$.
The lengths along the axis are in $\mu m$.
}
\label{fig6}
\end{figure}

Besides the stable equilibrium vortex position at 
interstitial sites of the triangular droplet lattice,
we have found that the saddle point between two adjacent droplets is a
(meta)stable equilibrium vortex position, with slightly 
higher energy than the interstitial equilibrium one. 
The map of the difference between the 
densities of the configuration
with and without vortex at the saddle point for the 
$a_s=93\,a_0$ case is shown in  Fig. \ref{fig6}.
Again, the density depression due to the vortex 
is very anisotropic and reflects the local symmetry of the system.

We also considered the possibility of a third equilibrium position 
with the vortex line piercing a droplet parallel to the 
polarization axis and passing through its center.
Such configuration has proven however to be unstable, and the vortex line 
is expelled during the imaginary time minimization,
ending in the lowest energy interstitial site configuration.
A similar unstable behavior has been observed for a vortex 
aligned with the polarization axis at
the center of self-bound dipolar droplets \cite{macri}.

\begin{figure}[t]
\centerline{\includegraphics[width=1.1\linewidth,clip]{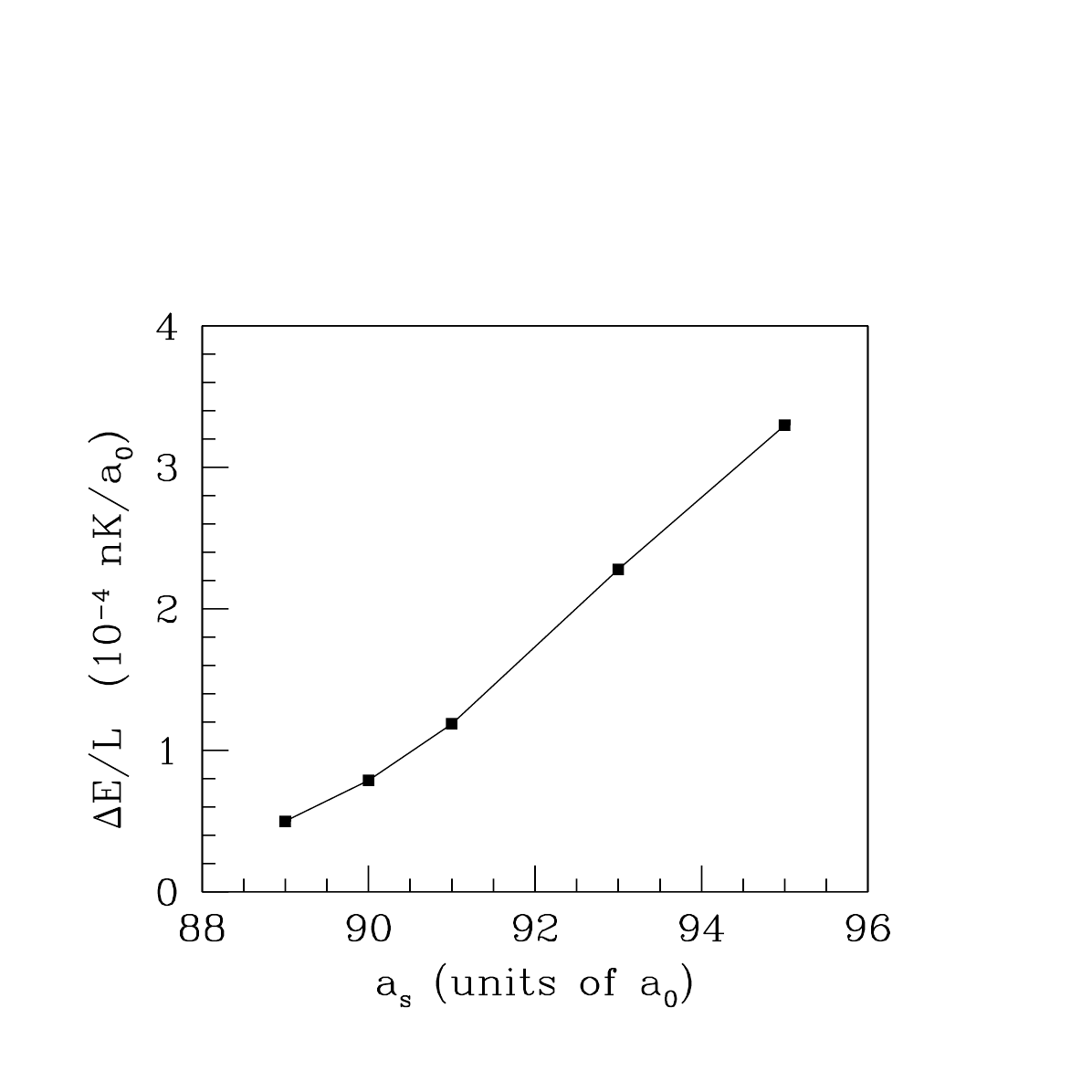}}
\caption{ Energy barrier per unit length in the $z$-direction 
between interstitial and saddle-point configurations
as a function of $a_s$.
}
\label{fig7}
\end{figure}

The difference between the energies of the saddle-point and interstitial site
configurations represents 
the barrier for displacing a vortex through the supersolid
lattice. The calculated energy barriers per unit length 
are shown in Fig. \ref{fig7} for different values of $a_s$.
Here $L$ in the length of the vortex line along the $z$-direction,
which we take as $L=\sqrt{\int \bar{\rho}(z) z^2 dz /\int \bar{\rho } (z) dz   }$,
where $\bar{\rho}(z)\equiv \int \rho (x,y,z)dx dy/(L_x L_y)$. We find
$L\sim 6.7\,\mu m$.
We notice the strong decrease of the energy 
barrier as the density of the superfluid background diminishes.  

\begin{figure}[t]
\centerline{\includegraphics[width=1.0\linewidth,clip]{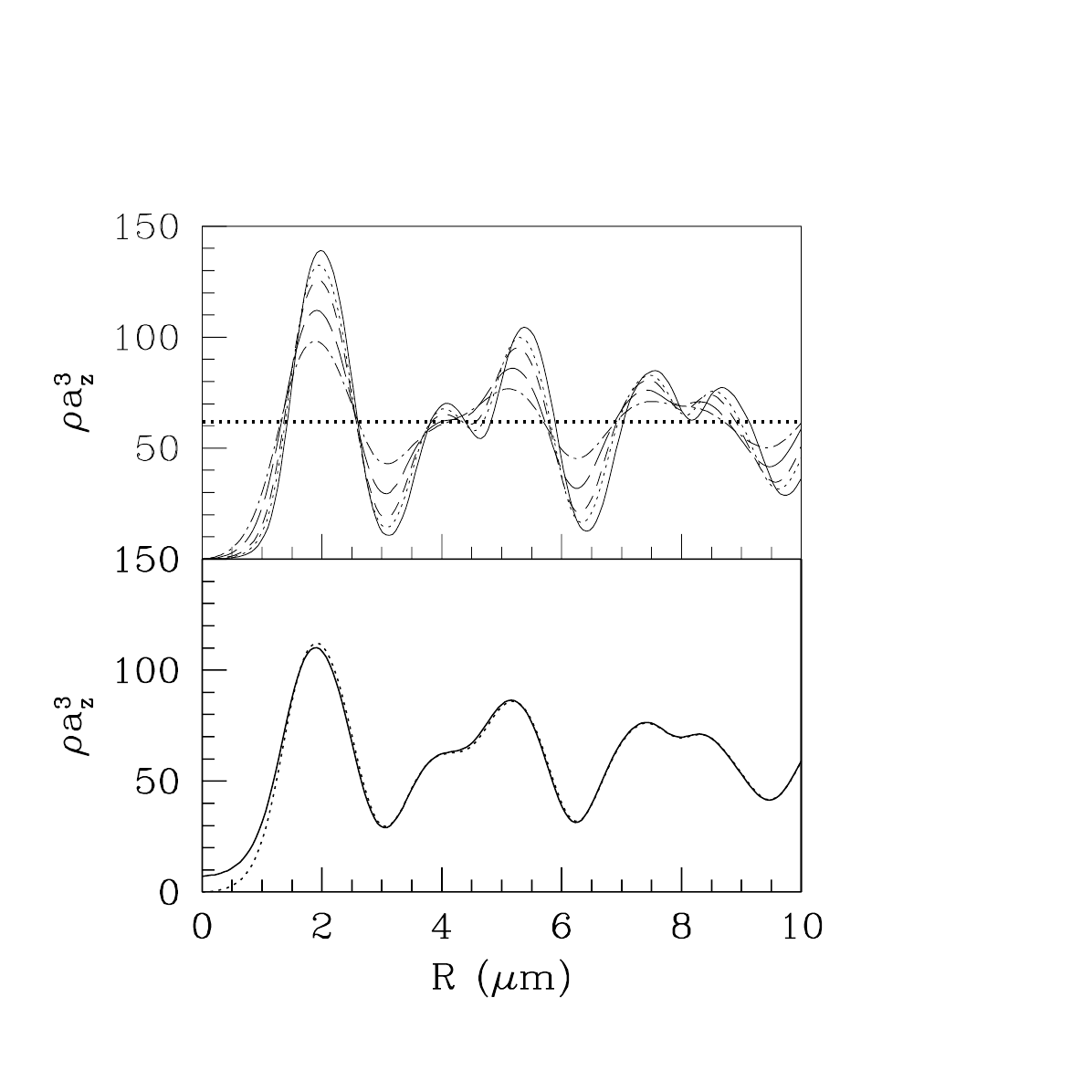}}
\caption{
Upper panel: angular average on the $z=0$ symmetry plane, of the number density around a vortex.
 From top to bottom --at the first maximum--
$a_s=89, 90, 91, 93$, and $95\,a_0$. The  nearly overlapping horizontal lines show the average densities
on the same plane.
Lower panel: angular average of the number density with and without vortex, for the $a_s=93\,a_0$ case.
}
\label{fig9}
\end{figure}

Although the very deformed and anisotropic shape of the vortex core in the SSP as
shown in Fig. \ref{fig4b} hinders a clear definition of the vortex core extension,
one can estimate it by performing  
an angular average of the density (on the $z=0$ symmetry plane) over the polar angle 
around the interstitial site where the vortex is placed.
We show this average in the lower panel of Fig. \ref{fig9}  
for the state with and without vortex corresponding to $a_s=93\,a_0$;
the upper panel  shows such angular average for 
different values of $a_s$. 
The horizontal lines show, in the same 
$z=0$ symmetry plane, the density values obtained by 
additional averaging over the radial distance, for the same $a_s$ values.
The radius $R_c$ of the averaged cavity is 
measured at half the value 
of the depression of the local density at the vortex axis, i.e.
from the condition 
$\rho _0(R_c)-\rho_v(R_c)=\rho_0 (0)/2$,
where $\rho _v$ and $\rho _0$ are the average density profiles 
with and without vortex, respectively.
The cavity radius so calculated ranges 
from $R_c\sim 23{,}000\,a_0$ for $a_s=95\,a_0$ to
$R_c\sim 35{,}000\,a_0$ for $a_s=89\,a_0$.
The calculated values are reported in Table I.
For comparison, 
the distance between the vortex core position and
the neighboring droplets center is $37{,}000\, a_0$ (for the $a_s=93\,a_0$ case).

A measure of the local vortex excitation energy per
unit length of a vortex of length $L$ is
the integrated vortex kinetic energy \cite{reatto,rossi}, which can be
defined as follows:

\begin{equation}
\epsilon_v(R) =\frac{1}{L} [E_{kin}^v(R)-E_{kin}(R)]
\end{equation}
where $E_{kin}^v,E_{kin}$ are the kinetic energies
within a cylinder of radius $R=\sqrt{x^2+y^2}$ and length $L$
(with and without a vortex line along its axis)
as a function of the distance $R$
from the vortex axis (here the $z$-axis).

Neglecting the $z$-dependence of the energy we compute
$\epsilon _v(R)$ by averaging 
the kinetic energy densities on the 
$z=0$ symmetry plane.
The kinetic energy per unit length is thus given by the integral
\begin{equation}
\frac {E_{kin}(R)}{L}=2\pi \int _0 ^R \epsilon (R')R' dR'
\end{equation}
where $\epsilon (R)$ is the 
angular average, performed in the $z=0$  plane, of 
$ (\hbar ^2/2m)|\nabla_R \phi|^2$, where 
$\nabla_R \equiv (\partial/\partial x,\partial/\partial y)$. Similar expressions hold for
$E^v_{kin}(R)/L$.

We notice that the classical hydrodynamical counterpart of $\epsilon _v(R)$ 
for a vortex of circulation $\kappa$ in an incompressible fluid of density $\rho _0$  is 

\begin{equation}
\epsilon _v ^{hydro}(R)=\frac {\kappa ^2}{4\pi} m 
\rho_0 \left[\ln\left(\frac{R}{d_v}\right)+\delta \right]
\label{hydro_class}
\end{equation}
where $d_v $ is the vortex core radius and $\delta$ 
depends on the core model  ($\delta=0$
for the hollow core model and $\delta=1/4$ for a 
core in rigid rotation) \cite{donnelly}. The parameter $\delta$ in the previous 
equation can be absorbed in the logarithmic 
term; using the quantum value for the circulation, $\kappa=h/m$,
for a singly-quantized vortex $\epsilon_v ^{hydro}$ reads

\begin{equation}
\epsilon _v ^{hydro}(R)=\frac{\hbar^2}{m}\pi 
\rho _0 \, \ln\left(\frac{R}{\lambda}\right)
\label{hydro}
\end{equation}
where $\lambda =d_ve^{-\delta}$ is the core parameter.

\begin{figure}[t]
\centerline{\includegraphics[width=1.0\linewidth,clip]{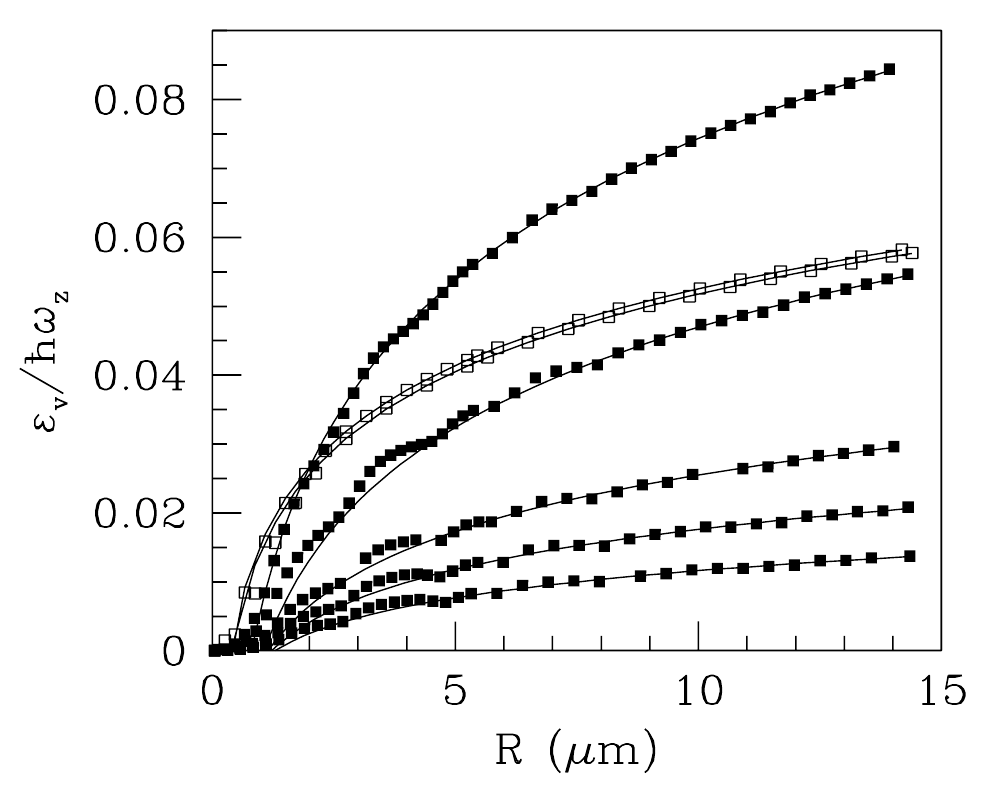}}
\caption{Integrated vortex kinetic energy $E_v(R)$ as a function of 
the distance from the vortex core. 
Filled squares, from top to bottom: $a_s=95,93,91,90,89\,a_0$ (SSP).
Open squares: $a_s=110, 100\,a_0$ (SFP).
The lines are the hydrodynamical form fit.
}
\label{fig10}
\end{figure}

We have used the above expression to fit the 
calculated $\epsilon _v(R)$ for the supersolid structures
investigated here with fitting parameters $\rho _0$ and $\lambda$.
We show in Fig. \ref{fig10}, for different values of $a_s$, the calculated $\epsilon_v(R)$ 
values (filled squares) 
and the best fit obtained by using the hydrodynamical approximation Eq. (\ref{hydro}) (solid line).
As expected, the approximation breaks down at distances approaching
the vortex core, as there the local density becomes very small.
One can notice from the figure that the computed 
kinetic energy has some weak oscillations whose
locations are related to the distances from the interstitial site where the vortex is placed, to 
the different droplet shells. 
For comparison, we also show in Fig. \ref{fig10}  
the results for the vortex-hosting SFP corresponding to two $a_s$ values (open squares).
Notice that the supersolid vortex kinetic energy at $a_s = 95\,a_0$ has
a higher value than in the superfluid phase.
We do not have a simple explanation for this jump.
The vortex kinetic energy 
in the supersolid differs from that in the superfluid 
for the very different character of the density profile and 
for a flow pattern much different from the tangential 
flow of the superfluid. We have not tried to disentangle the two effects.

We report in Table I the calculated values for 
the core parameter $\lambda$ and prefactor $\rho _{0}$
as obtained from the fit.
At variance with the SFP, for which the calculated value of $\rho _{0}$
coincides with the average density on the $z=0$ plane (4th column 
in the Table), it does not for the SSP: here the
calculated $\rho _{0}$ values show a marked decrease with $a_s$.
One may argue that these values should be related to the superfluid halo 
embedding the droplets, and thus to 
the (minimum) density in the interstitial regions
between droplets (5th column in the Table): 
this is not the case though, since the minimum density values
decrease with $a_s$ much faster than $\rho _{0}$.
We also show in the Table the vortex
``core'' energy, defined as $E _v(R_c)$, where the core radius $R_c$
has been estimated as described before.

The core parameter $\lambda$ and core energy $E_c$ 
depend only weakly on the density in the SFP whereas 
the core radius $R_c$ increases as $a_s$ decreases, 
i.e. as the dipolar interaction becomes more important. 
In the SSP there is a clear change in the dependence of 
these vortex parameters as a function of $a_s$. 
As mentioned before, $\rho_0$ is a strongly decreasing 
function of $a_s$. Both $\lambda$ and $R_c$ 
increase significantly for decreasing $a_s$, and 
this means that the vortex size expand as the 
value of the low density halo decreases. 
The core energy has the opposite dependence on $a_s$, i.e. 
$E_c$ decreases as $a_s$ does, 
indicating that the decrease of the halo 
density has a stronger effect on the $E_c$ value than 
the increase of the vortex core radius.

\begin{table*}[!]
  \begin{center}
    \label{tab:table1}
        \begin{tabular}{ccccccccccccc} 
\hline
\hline    
      $a_s \,(a_0)$ &$\quad$&$\rho _{0}\,(a_0^{-3}) $ &$\quad$& $\lambda \,(a_0)$ &$\quad$& $\rho _{av}\,(a_0^{-3})$ &$\quad$& $\rho _{min}\,(a_0^{-3})$ &$\quad$& $R_c\,(a_0)$ 
      &$\quad$& $E_c$  (nK/$a_0)$ \\
      \hline
  89\,(SSP) & $\quad$&$1.23\times 10^{-11}$ &$\quad$& 24{,}360 &$\quad$& $3.45\times 10^{-11}$ &$\quad$& $4.36\times 10^{-13}$ &$\quad$& $35{,}221$ 
  &$\quad$& $2.27\times 10^{-2}$\\
  90\,(SSP)  &$\quad$&$1.82\times 10^{-11}$ &$\quad$& 23{,}160 &$\quad$& $3.46\times 10^{-11}$ &$\quad$& $7.97\times 10^{-13}$ &$\quad$& $32{,}704$ 
  &$\quad$& $3.18\times 10^{-2}$\\
  91\,(SSP)  &$\quad$&$2.56\times 10^{-11}$ & $\quad$ &21{,}790 &$\quad$& $3.47\times 10^{-11}$ &$\quad$& $1.40\times 10^{-12}$ &$\quad$& $30{,}474$ &$\quad$& $4.33\times 10^{-2}$ \\
  93\,(SSP) &$\quad$& $4.56\times 10^{-11}$ &$\quad$& 20{,}180 &$\quad$& $3.49\times 10^{-11}$ &$\quad$& $4.01\times 10^{-12}$ &$\quad$& $26{,}130$ 
  &$\quad$& $6.50\times 10^{-2}$\\
  95\,(SSP) &$\quad$& $6.44\times 10^{-11}$ &$\quad$& 15{,}420 &$\quad$& $3.51\times 10^{-11}$ &$\quad$& $1.09\times 10^{-11}$ &$\quad$& $22{,}644$ 
  &$\quad$& $8.40\times 10^{-2}$\\
  \hline
 98\,(SFP) &$\quad$& $3.54\times 10^{-11}$ &$\quad$& 7{,}484  &$\quad$& $3.53\times 10^{-11}$ &$\quad$& ---  &$\quad$&$17{,}755$ &$\quad$& $5.56\times 10^{-2}$\\
 100\,(SFP) &$\quad$& $3.52\times 10^{-11}$ &$\quad$& 7{,}777  &$\quad$& $3.52\times 10^{-11}$ &$\quad$& ---  &$\quad$&$16{,}470$ &$\quad$& $5.72\times 10^{-2}$\\
 105\,(SFP) &$\quad$& $3.50\times 10^{-11}$ &$\quad$& 7{,}612  &$\quad$& $3.50\times 10^{-11}$ &$\quad$& --- &$\quad$& $14{,}350$ &$\quad$& $5.97\times 10^{-2}$\\
 110\,(SFP) &$\quad$& $3.48\times 10^{-11}$ &$\quad$& 7{,}200  &$\quad$& $3.48\times 10^{-11}$ &$\quad$& --- &$\quad$& $12{,}750$ &$\quad$& $6.03\times 10^{-2}$\\
 \hline
 \hline
    \end{tabular}
        \caption{
Parameters $\rho_0$ and $\lambda$ of the fit of the vortex kinetic energy 
with the hydrodynamic form Eq. (\ref{hydro}) for some values of $a_s$ 
in the superfluid (SFP) and in the supersolid (SSP). 
Other reported quantities are the average density 
$\rho_{av}$ on the $z=0$ plane, the density $\rho_{min}$ 
at the interstitial site of the supersolid, the radius $R_c$ 
of the vortex cavity, and the vortex core energy $E_c$.
}
  \end{center}
\end{table*}

An important issue is the nature of the 
low energy excitations of a vortex in the SSP. 
We already mentioned a difference between 
a vortex in a superfluid and in a supersolid: 
in a homogeneous superfluid a vortex can move 
without restriction and its low energy 
excitations are Kelvin waves whose 
energy in an unbounded fluid vanishes as the wave 
vector $k$ goes to zero \cite{lamb,donnelly}. 
We have shown that in a supersolid a vortex has a discrete set
of equilibrium positions and an energy
barrier is present for its motion from
one site to another. 
Therefore, the vortex is a localized excitation 
in the plane perpendicular to the
vortex axis and this should modify the properties of 
the Kelvin waves. In particular their spectrum should 
display an energy gap because the vanishing of 
the spectrum of the Kelvin waves in the limit of zero 
wave vector is a consequence of the translational 
invariance of the vortex excitation in a uniform superfluid, 
invariance that is absent in the supersolid. 
This predicted gap in the supersolid has a completely 
different origin from the small energy 
gap of Kelvin waves when bosons are in a trap \cite{fetter1}.

There is the possibility that the 
low energy excitations of a vortex 
in a supersolid have a quite different character
when the droplets are very elongated
along the $z$-axis. Such excitations 
could be kinks along the vortex 
axis such that the vortex core moves 
from one equilibrium site to another site along the axis. 
The energy of the vortex 
in a SSP is a periodic function of position 
and this plays a role similar to  the 
Peierls potential acting on a dislocation 
in a crystalline solid. 
Kinks are indeed the low energy 
excitations of a dislocation in solid $^4$He \cite{beamish}.
The study of the excitations of a 
vortex in a supersolid is an interesting 
problem but it is quite outside the scope of the present paper.

\section{Vortex dipole in dipolar bosons}

As discussed in Sect. IV, vortices in a dipolar supersolid 
have well defined equilibrium positions in the 
``solid'' structure so that the vortex energy  is a periodic function of its position.
Therefore, a number of properties are expected to differ 
from those in a superfluid. 
In this Section we study one of them, namely the dynamics of 
a vortex dipole, i.e. two counter-rotating straight parallel vortices.
Vortex dipoles have been created and observed in BECs confined 
by parabolic potentials \cite{dipo_bec}, and
the motion of vortex dipoles in superfluid $^4$He has been 
simulated within the time-dependent DFT approach \cite{dipole}.

In classical hydrodynamics of incompressible fluids \cite{lamb}
a vortex dipole is a stable entity that moves with a constant 
velocity that is perpendicular to the plane containing the 
vortices and inversely proportional 
to the distance $d$ between them. 
The same behavior holds in superfluid systems
when $d$ is much larger than the healing length
(which in the present case can be estimated from the 
vortex core sizes in Fig. \ref{fig2-SM});
in the superfluid,
the vortex dipole propagates with a constant 
velocity $v_d=\hbar/(m\,d)$ \cite{donnelly}. 

We have addressed the dynamics of a vortex dipole in the extended 
dipolar bosons system. To this end, we have first  
prepared the initial vortex dipole configuration. This
can be done generalizing as follows Eq. (\ref{imprint}) in Sec. II:
\begin{eqnarray}
\phi_0(\mathbf{r})&=&\rho_0^{1/2}(\mathbf{r}) \left[ {(x-x_v)+\imath (y-y_v) \over \sqrt{(x-x_v)^2+(y-y_v)^2}}  \right.
\nonumber
\\
&&\left.{(x-x_{\bar{v}})-\imath (y-y_{\bar{v}}) \over \sqrt{(x-x_{\bar{v}})^2+(y-y_{\bar{v}})^2}} 
 \right]
\label{imprintdipole}
\end{eqnarray}
where $(x_v, y_v)=(d/2, 0)$ and $(x_{\bar{v}}, y_{\bar{v}})=(-d/2,0)$
We evolve this wave function in 
imaginary time until a stationary configuration is reached. Since 
the total circulation is zero, usual periodic boundary conditions 
can be used. 

It should be noticed that the dipole wave function is not 
orthogonal to that of the ground state. Hence, a very long 
imaginary time evolution would bring the 
initial state $\phi_0(\mathbf{r})$ to the ground 
state of the system, i.e. the vortex dipole would disappear.
However, 
 we find that after a relatively short imaginary time
$\tau_c$ the initial phase settles down to a quasi-stationary state 
where the vortex cores are fully developed while remaining in their
initial positions
(whereas orders of magnitude longer imaginary times would be 
needed to recover the vortex-free ground state). 
At this point we initiate the evolution  of the wave function in real time.
By using different values of $\tau_c$ we have verified
that the real time evolution does not depend on the used  $\tau_c$  value.

\begin{figure}[t]
\centerline{\includegraphics[width=1.05\linewidth,clip]{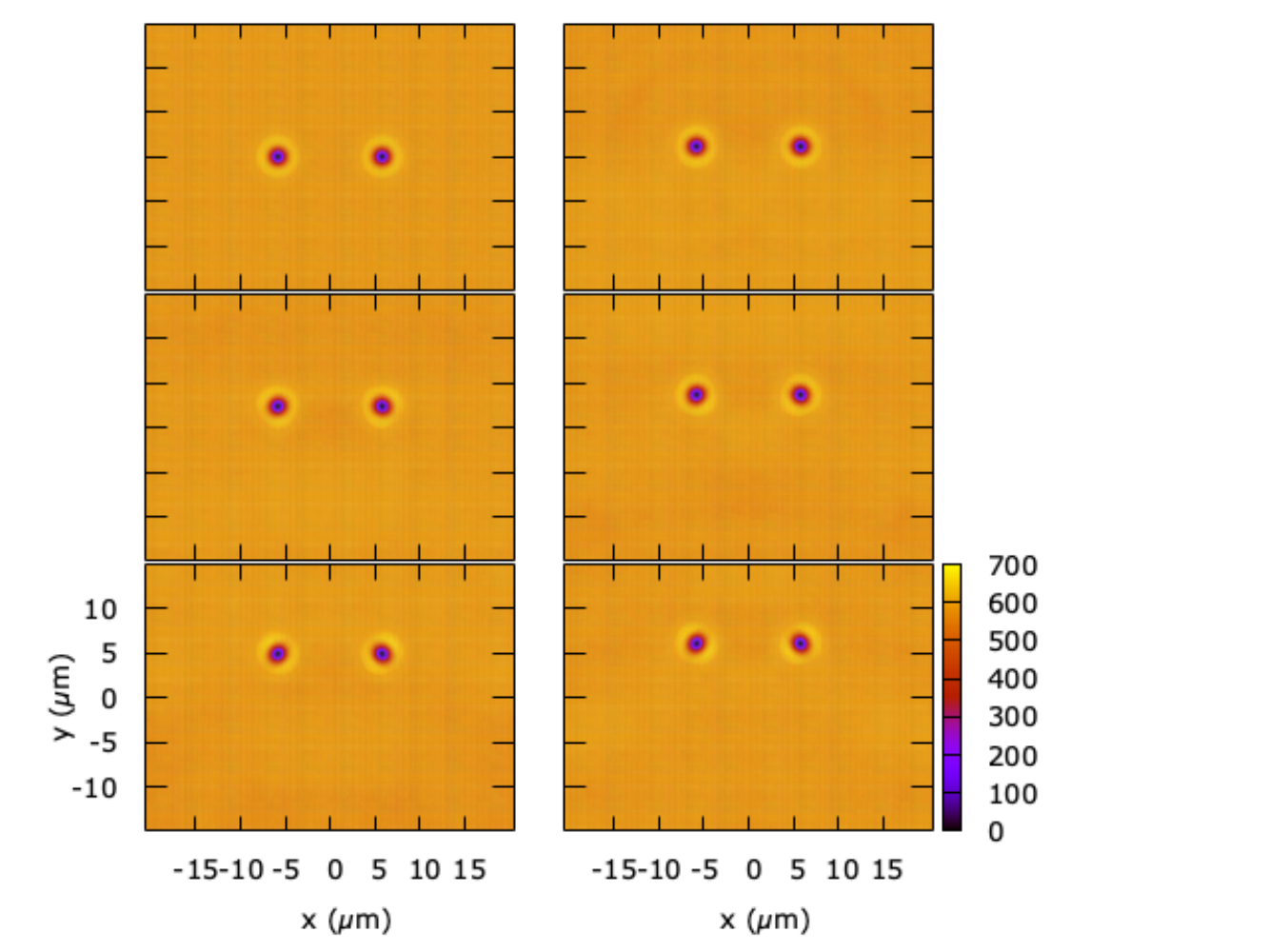}}
\caption{Vortex-antivortex dynamics in real time,
for the case $a_s=110\,a_0$ (SFP). The density, integrated along the $z$-axis,
is shown in units of $a_{ho}^{-2}$.
Snapshots are taken, from top to bottom and from left to right,
at $t=0, 24, 48, 72, 96$, and $120$\,ms.
Lengths are in $\mu$m.
}
\label{fig12}
\end{figure}

We have first studied the vortex dipole in the SFP. In this case we 
find that the vortex dipole translates 
with a constant velocity, and this velocity is of order 
of that given by classical hydrodynamics,
$v_{d}=\hbar/(m\,d)$.
Notice that the observed velocity 
can differ from the hydrodynamic result for
two reasons, because $d$ is not very large compared 
to the healing length and because the periodic 
boundary conditions modify the streamlines 
of the dipole near the cell boundaries
with respect to those in an unbounded fluid \cite{roberts}.

The real time evolution of a vortex dipole in the SFP for $a_s=110\,a_0$ is shown in Fig. \ref{fig12},
in the form of snapshots  of the density taken at increasing times during the dynamics. 
The vortex dipole has been prepared with an initial 
vortex-antivortex distance $d=6.5\,\mu m$.

\begin{figure}[t]
\centerline{\includegraphics[width=1.05\linewidth,clip]{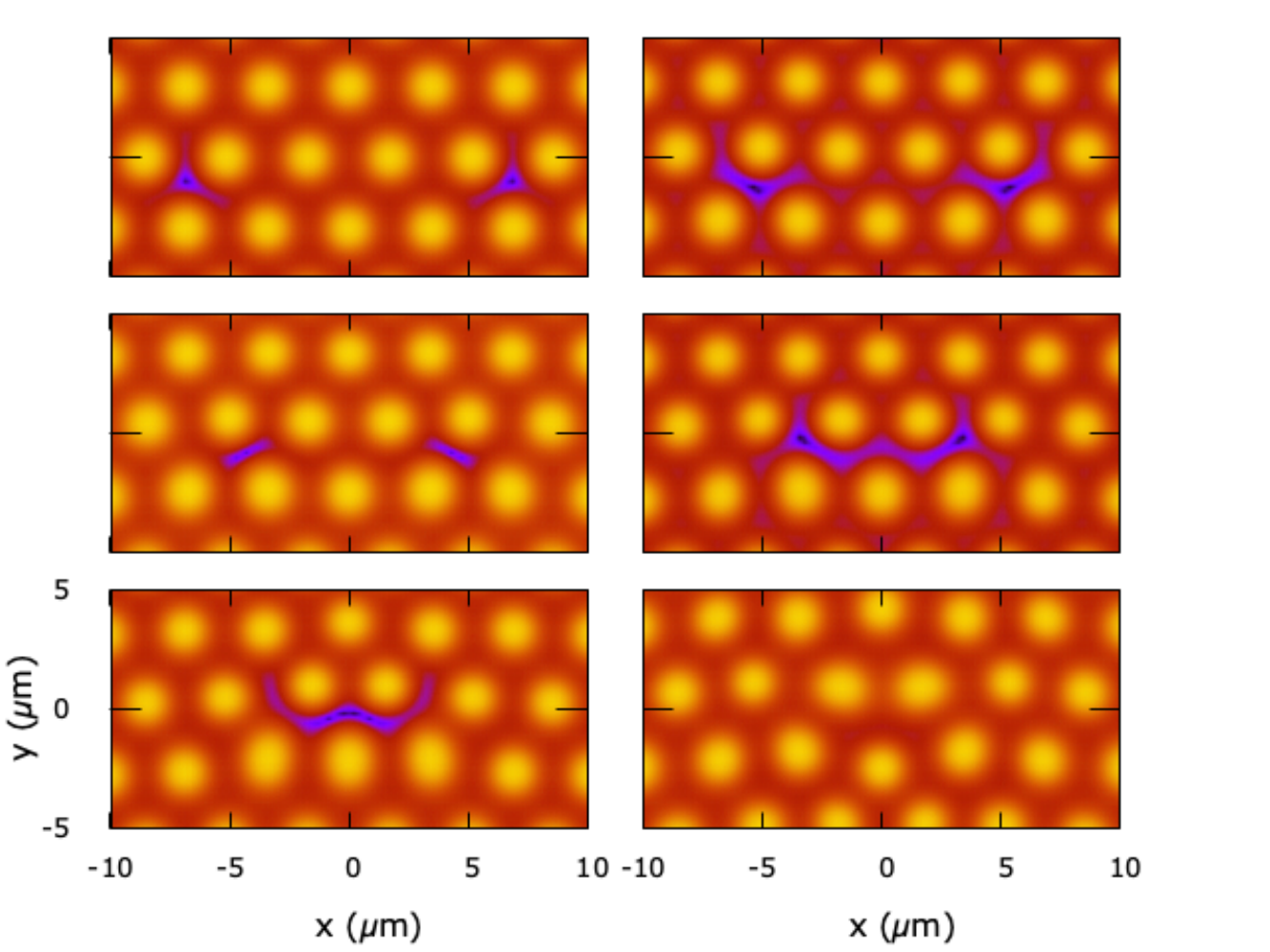}}
\caption{Vortex-antivortex dynamics in real time,
for the case $a_s=93\,a_0$ (SSP). For clarity, the logarithm of the $z$-averaged density
(in units of $a_{ho}^{-2}$) is shown.
Snapshots are taken, from top to bottom and from left to right,
at $t=0, 27, 36, 41, 48$, and $75$\,ms.
Lengths are in $\mu$m.
}
\label{fig13}
\end{figure}

A completely different behavior is found in the SSP: no 
translation of the dipole takes place, and 
the two vortices approach until they annihilate 
each other. The dynamics of a vortex dipole in the SSP 
is displayed in Fig. \ref{fig13}, where it appears that the 
vortex-antivortex pair undergo annihilation in a very short time.
Notice that a logarithmic scale is used in this figure
in order to highlight the vortex positions, which would
be otherwise invisible in an ordinary plot.
The results presented in Fig. \ref{fig13} 
are for $a_s=93\,a_0$ and initial intervortex distance 
$d=4a$, where $a=3.44\,\mu m$ is the lattice parameter of the SSP for this $a_s$ value. 
We also considered a different, more symmmetric initial configuration of
the two vortices, i.e. the case where the
two vortices are positioned along the $y$-axis of  
Fig. \ref{fig13} and separated directly by a droplet.
In spite of the
symmetric initial configuration of the two vortices, which 
might suggest the possibility
of a metastable state where the two vortices remain pinned there,  
the combined velocity field starts displacing both vortices to the
right (it would be the left for interchanged vortex signs), 
and then they almost immediately annihilate each other.
We have found that similar dynamics of the vortex dipole occur 
for other values of $d$ and $a_s$ corresponding to a SSP.
Therefore, the disclosed annihilation process of 
two counter-rotating vortices appears to be generic 
for the SSP of dipolar bosons.

By following the snapshots in Fig. \ref{fig13}
one can see that the recombination process proceeds by a set of jumps 
from one interstitial site to a neighboring one and that in the jump 
process each vortex becomes elongated by extending over the two 
neighboring sites as well as over the in-between saddle point. 
Since the snapshots represent the projected density on
the $x-y$ plane one can ask if the elongation of the vortex during
the jump is due to propagation of kinks along the vortex axis
as discussed in the previous Section or if the vortex rigidly
translates from one site to another one. We find that this
second case indeed occurs since we have verified that the density 
in single $x-y$ planes 
has exactly the same dynamics of the projected one.

\begin{figure}[t]
\centerline{\includegraphics[width=1.0\linewidth,clip]{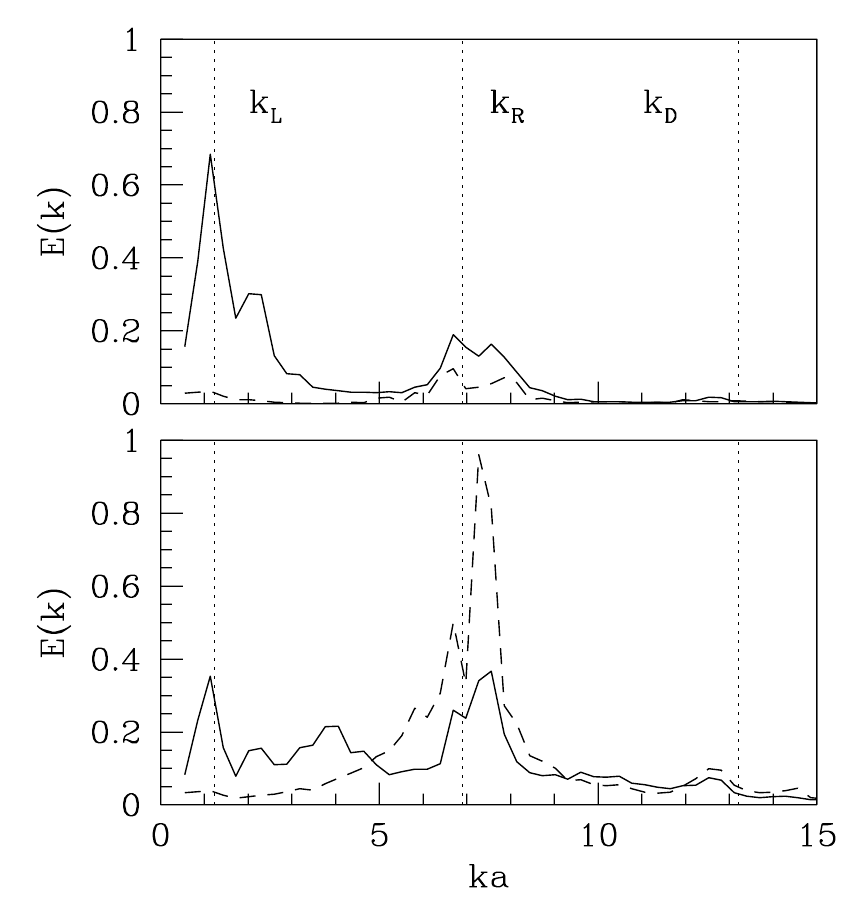}}
\caption{Energy spectrum (in units of $1\times 10^{-18}$ hartree $\times a_0$) 
before (top) and after (bottom) the vortex dipole
annihilation in the SSP with $a_s=93\,a_0$.
Solid lines, incompressible spectrum; dashed lines, compressible spectrum. 
The meaning of the vertical lines is explained in the text.
}
\label{fig18}
\end{figure}

The energy released immediately after the annihilation goes into excitations 
of the droplet array, which appears to be deformed 
and oscillating after the recombination event. 
Such change in the character of the excitation energy can be characterized by computing 
the spectral density of the kinetic energy  of 
the superfluid velocity field, decomposing it into compressible 
and incompressible parts \cite{Nor97,tsubota} by utilizing the
Helmholtz decomposition theorem.
We show in Fig. \ref{fig18} these 
two contributions as a function of the wavevector in the $x-y$ 
plane $k$, where we have taken a time average 
over the first $40\,ms$ of the dynamics, which corresponds to times before  
vortex annihilation, and over $20\,ms$ after vortex 
annihilation. 
The vertical lines show three relevant wave vectors:
$k_L=2\pi/L$ where $L=\sqrt{L_xL_y}$ is the average size
of the simulation cell, $k_R=2\pi/a$ is the 
wave vector corresponding to the periodic arrangement of droplets 
($a$ being the droplet-droplet nearest-neighbor distance 
in the SSP), and $k_D=2\pi/D$ where $D$ is the average droplet diameter.
One can see that when the vortex dipole is present (first time interval),
the main contribution to the spectral density is in the 
incompressible component at rather small $k$. 
In particular, the peak at $k_L$ is related to the cell size
through the periodic boundary conditions. 
After the vortex dipole annihilation, the main contribution is
transferred into the compressible component at 
$k$ values $\sim k_R$ which
corresponds to the 
$k$ value associated with 
solid order. A similar
roton burst has been observed in the annihilation of vortex dipoles in 
superfluid $^4$He nanodroplets \cite{escartin}.

\section{Summary}

It is now well established that dipolar bosons, in addition 
to a superfluid state, can be in a supersolid state 
in which the local density of the condensate is modulated in the plane
perpendicular to the polarization direction. 
This phase appears when the strength of the dipolar 
interaction is strong enough compared to 
the contact interaction. The supersolid state 
consists of a regular array of droplets and a 
low-density superfluid background which occupies the space between them.

In order to determine the intrinsic 
properties of vortices in the SSP we have studied 
the system without trapping potential in the $x-y$ plane 
but with suitable boundary conditions in the computation 
cell, and a harmonic trap is present only in the $z$-direction, 
the polarization direction.
We have determined the equilibrium lattice parameter and 
the superfluid fraction of this extended system of dipolar bosons
for different values of the s-wave scattering length.
We have found that a singly quantized, linear vortex 
along the polarization direction is a stable 
excitation of the system in a stationary state. 
A fundamental property of a vortex in the supersolid 
state is that it has equilibrium positions in 
the crystalline structure of the bosons, in our 
case at the interstitial sites of the triangular droplets lattice. 

We have characterized the density profile of the vortex, its velocity field and the 
kinetic energy as function of the distance from the vortex core. 
The velocity field strongly 
differs from the simple circular
flow in a superfluid. Besides,
most of the contribution to the quantum of circulation in the SSP
is due to the low-density background, the droplets 
giving only a small contribution to it. 

As the ratio of the dipolar to the contact interaction increases,
the vortex core expands and the core energy decreases. 
We have computed the energy barrier that a 
vortex has to surmount to move from an equilibrium site to a neighboring 
one. 
The existence of this energy barrier -and the density modulation 
due to the presence of droplets– represents a major difference 
from a uniform superfluid in which a vortex is free to move 
when it is far from the boundaries. The dynamics of vortices 
in a dipolar superfluid has been predicted \cite{mulker} 
to differs from that in a standard BEC. 
Even stronger modifications of the vortex dynamics 
are expected in a supersolid. 
In a SSP such dynamics should be characterized by 
oscillations around the equilibrium position and 
by jumps from one site to another.

We have explored one particular aspect of the 
vortex dynamics, namely the behavior of a vortex dipole made of
two counter-rotating parallel vortices. In a uniform superfluid 
the vortex dipole is a stable entity which rigidly 
translates with constant velocity. We have found that in the SSP
the vortex dipole is an unstable excitation, 
the two vortices of opposite chirality do not translate but 
they approach each other by a series of jumps until they 
annihilate and the excitation energy goes into bulk 
excitation of the supersolid. We believe that this is 
just an example of new features of the vortex dynamics  
in a supersolid. For instance Kelvin waves, the basic 
excitations of a vortex, should be strongly modified in the SSP with 
respect to the SFP and a new kind of excitation 
might arise, at least when the droplets are elongated enough, 
consisting of kinks of the vortex core transferring the 
core of the vortex from one equilibrium site to another.

Finally, it should also be noticed that the periodic 
boundary conditions in the plane perpendicular to the polarization direction
impose a constraint on the spatial order
of the droplets, so that they cannot rotate even
when a vortex is present. However, we believe that this  
is not an artifact of boundary conditions.
In fact, we have verified that a single vortex is 
also a (meta)stable excitation when the system is 
confined in the $x-y$ plane by a cylindrically symmetric trap. 
We find that the droplets do not rotate either during the typical 
lengths of our simulation, of order of several tens of milliseconds.
This gives evidence for the existence of an excitation 
with the vortex phase singularity at the center of 
the bucket with a stationary supersolid structure.
This behavior is at variance with the state of 
dipolar bosons in a rotating trap which induces 
single or multiple vortices \cite{stringa19,stringa20,arxiv}
with the whole supersolid structure set in rotation.
It will be interesting to study the behavior of a
vortex dipole for other systems in a SSP, especially 
if the spatial order is different from that of dipolar
bosons.

\begin{acknowledgments}
One of us (F.A.) is indebted to Elena Poli for 
contributing to the early stages of this work. 
This work has been supported by Grant No.  FIS2017-87801-P (AEI/FEDER, UE) (M.B., M.P.).
\end{acknowledgments}


\begin{thebibliography}{99}

\bibitem{donnelly} 
R.J. Donnelly, {\it Quantized vortices in helium II},  Cambridge University Press (1991).

\bibitem{barenghi} 
C.F. Barenghi, R.J. Donnelly and W.F. Vinen, {\it Quantized Vortex Dynamics and Superfluid Turbulence}
(Springer Science and Business Media, Berlin, 2001).

\bibitem{fetter} 
A.L. Fetter, Rev. Mod. Phys. {\bf 81}, 647 (2009).

\bibitem{Pit16}
L. Pitaevskii and S. Stringari,  {\it Bose-Einstein Condensation and Superfluidity}, International Series of
Monographs on Physics vol. 164 Oxford University Press, U.K.  (2016).

\bibitem{bal10} 
S. Balibar, Nature {\bf 464}, 176 (2010).

\bibitem{and69} 
A.F. Andreev and I.M. Lifshitz, Sov. Phys. JETP {\bf 29}, 1107 (1969).

\bibitem{che70} 
G.V. Chester, in {\it Lectures in Theoretical Physics}, ed. by 
K.T. Mahanthappa and W.E. Britten (Gordon \& Breach, New York, 1969); 
{\it Lecture notes of the Summer Institute for Theoretical Physics}, 1968, Boulder (CO, USA); 
G.V. Chester, Phys. Rev. A {\bf 2}, 256 (1970).

\bibitem{leg70} 
A.J. Leggett, Phys. Rev. Lett. {\bf 25}, 1543 (1970).

\bibitem{cha13} 
M.H.W. Chan, R.B. Hallock, and L. Reatto, J. Low Temp. Phys. {\bf 172}, 317 (2013); J. Low Temp. Phys. {\bf 173}, 354(E) (2013).

\bibitem{bon12} 
M. Boninsegni and N.V. Prokof'ev, Rev. Mod. Phys. {\bf 84}, 759 (2012).

\bibitem{lamb} 
H. Lamb, {\it Hydrodynamics} (Dover Publications, Inc., New York, 1945).

\bibitem{lewe} 
L. Santos, G.V. Shlyapnikov, and M. Lewenstein, Phys. Rev. Lett. {\bf 90}, 250403 (2003).

\bibitem{odell} 
D.H.J. O'Dell, S. Giovanazzi, and G. Kurizki, Phys. Rev. Lett. {\bf 90}, 110402 (2003).

\bibitem{wenzel} 
M. Wenzel, F. Bottcher, T. Langen, I. Ferrier-Barbut, and T. Pfau, Phys. Rev. A {\bf 96}, 053630 (2017).

\bibitem{blackie_baillie3} 
D. Baillie and P.B. Blakie, Phys. Rev. Lett. {\bf 121}, 195301 (2018).

\bibitem{Chomaz2018} 
L. Chomaz, R.M.W. van Bijnen, D. Petter, G. Faraoni, S. Baier, J.H. Becher, M.J. Mark, F. W\"achtler, L. Santos, and 
F. Ferlaino, Nature Physics {\bf 14}, 442-446 (2018).

\bibitem{kora} 
Y. Kora and M. Boninsegni,  J. Low Temp. Phys. {\bf 197}, 337 (2019).

\bibitem{ancilotto} 
S.M. Roccuzzo and F. Ancilotto, Phys. Rev. A {\bf 99}, 041601(R) (2019).

\bibitem{Kad16} 
H. Kadau, M. Schmitt, M. Wenzel, C. Wink, T. Maier, I. Ferrier-Barbut, and T. Pfau, Nature {\bf 530}, 194 (2016).

\bibitem{Fer16}
I. Ferrier-Barbut, H. Kadau, M. Schmitt, M. Wenzel, and T. Pfau, Phys. Rev. Lett.  {\bf 116}, 215301 (2016).

\bibitem{schmitt} 
M. Schmitt, M. Wenzel, F. B\"ottcher, I. Ferrier-Barbut, and T. Pfau, Nature {\bf 539}, 259 (2016).

\bibitem{santos_wachtler}  
F. W\"achtler and L. Santos, Phys. Rev. A {\bf 93}, 061603(R) (2016).

\bibitem{blackie_baillie1} 
D. Baillie, R.M. Wilson, R.N. Bisset, and P.B. Blakie, Phys. Rev. A {\bf 94}, 021602(R) (2016).

\bibitem{ferlaino} 
L. Chomaz, S. Baier, D. Petter, M.J. Mark, F. Wachtler, L. Santos, and F. Ferlaino, Phys. Rev. X {\bf 6}, 041039 (2016).

\bibitem{lima_pelster} 
A.R.P. Lima and A. Pelster, Phys. Rev. A {\bf 84}, 041604(R) (2011).

\bibitem{bombin} 
R. Bombin, J. Boronat, and F. Mazzanti, Phys. Rev. Lett. {\bf 119}, 250402 (2017).

\bibitem{cinti1} 
F. Cinti and M. Boninsegni, Phys. Rev. A {\bf 96}, 013627 (2017).

\bibitem{pohl} 
Y.-C. Zhang, F. Maucher, and T. Pohl, Phys. Rev. Lett. {\bf 123}, 015301 (2019).

\bibitem{chomx}
L. Chomaz, D. Petter, P. Ilzh\"ofer, G. Natale, A. Trautmann, C. Politi, G. Durastante, R.M.W. van Bijnen, A. Patscheider, M. Sohmen,
M.J. Mark, and F. Ferlaino, Phys. Rev. X {\bf 9}, 021012 (2019). 

\bibitem{bott19}
F. B\"ottcher, J.-N. Schmidt, M. Wenzel, J. Hertkorn, M. Guo, T. Langen, and Tilman Pfau, Phys. Rev. X {\bf 9},  011051 (2019).

\bibitem{tanzi} 
L. Tanzi, E. Lucioni, F. Fama, J. Catani, A. Fioretti, C. Gabbanini, R.N. Bisset, L.Santos and G. Modugno, Phys. Rev. Lett. {\bf 122}, 130405 (2019).

\bibitem{tanzi_new} 
L. Tanzi, S. M. Roccuzzo, E. Lucioni, F. Fama, A. Fioretti, C. Gabbanini, G. Modugno, A. Recati, and S. Stringari, 
Nature {\bf 574}, 382 (2019).

\bibitem{tanzi19} 
L. Tanzi, J.G. Maloberti, G. Biagioni, A. Fioretti, G. Gabbanini, and G. Modugno, arXiv:1912.01910 (2019).

\bibitem{stringa19} 
S.M. Roccuzzo, A. Gallemi, A. Recati, and S. Stringari, Phys. Rev. Lett. {\bf 124}, 045702 (2020).

\bibitem{ferla19}
P. Ilzh\"ofer, M. Sohmen, G. Durastante, C. Politi, A. Trautmann,
G. Morpurgo, T. Giamarchi, L. Chomaz, M. J. Mark, and F. Ferlaino, arXiv:1912.10892 (2019).

\bibitem{sep_joss_rica}
Y. Pomeau and S. Rica, Phys. Rev. Lett. {\bf 72}, 2426 (1994);
N. Sepulveda, C. Josserand, and S. Rica, Eur. Phys. J. B {\bf 78}, 439 (2010).

\bibitem{pfau} M. Guo, F. B\"ottcher, J. Hertkorn, J-N. Schmidt, M. Wenzel, H.P. B\"uchler,
T. Langen, and T. Pfau, Nature {\bf 574}, 386 (2019).

\bibitem{stringa20} 
A. Gallemi, S.M. Roccuzzo, S. Stringari and A. Recati, Phys. Rev. A {\bf 102}, 023322 (2020).

\bibitem{arxiv}
F. Ancilotto, M. Barranco, M. Pi, and L. Reatto, arXiv:2002.05934.

\bibitem{roberts} 
C.A. Jones and P.H. Roberts, J. Phys. A {\bf 15}, 2599 (1982); A. Griffin, 
V. Shukla, M.-E. Brachet, and S. Nazarenko, Phys. Rev. A {\bf 101}, 053601 (2020).

\bibitem{Anc17}
F. Ancilotto, M.  Barranco, F. Coppens, J. Eloranta, N. Halberstadt, A. Hernando, D. Mateo, and M. Pi,
Int. Rev. Phys. Chem. {\bf 36}, 621 (2017).

\bibitem{lahaye} 
T. Lahaye, C. Menotti, L. Santos, M. Lewenstein, and T. Pfau, Rep. Progr. Phys. {\bf 72}, 126401 (2009).

\bibitem{Pi07}
M. Pi, R. Mayol, A. Hernando, M. Barranco, and F. Ancilotto, J. Chem. Phys. {\bf 126}, 244502 (2007).

\bibitem{sadd} 
M. Sadd, G. V. Chester, and L. Reatto, Phys. Rev. Lett. 79, 2490 (1997).

\bibitem{Anc18}
F. Ancilotto, M. Barranco and M. Pi, Phys. Rev. B {\bf 97}, 184515  (2018).

\bibitem{Yi06}
S. Yi and H. Pu, Phys. Rev. A {\bf 73}, 061602(R) (2006).

\bibitem{reatto}
I. Amelio, D.E. Galli, and L. Reatto, Phys. Rev. Lett. {\bf 121}, 015302 (2018).

\bibitem{rocc} 
S.M. Roccuzzo, A. Gallemi, A. Recati, and S. Stringari, Phys. Rev. Lett. {\bf 124}, 045702 (2020).

\bibitem{macri}
A. Cidrim, F.E.A. dos Santos, E.A.L. Henn and T. Macri, Phys. Rev. A {\bf 98}, 023618 (2018).

\bibitem{rossi}
D.E. Galli, L. Reatto, and M. Rossi, Phys. Rev. B {\bf 89}, 224516 (2014).

\bibitem{fetter1} A.L. Fetter, Phys. Rev. A {\bf 69}, 043617 (2004). 

\bibitem{beamish} 
J. Beamish and S. Balibar, Rev. Mod. Phys. {\bf 92}, 045002 (2020)]. 

\bibitem{dipo_bec} 
T.W. Neely, E.C. Samson, A.S. Bradley, M.J. Davis, and B.P. Anderson, Phys. Rev. Lett. {\bf 104}, 160401 (2010);
D.V. Freilich, D.M. Bianchi, A.M. Kaufman, T.K. Langin, and D.S. Hall, Science {\bf 329}, 1182 (2010).

\bibitem{dipole} 
F. Ancilotto, M. Barranco, J. Eloranta, and M. Pi, Phys. Rev. B {\bf 96}, 064503 (2017).

\bibitem{Nor97}
C. Nore, M. Abid, and M.E. Brachet, Phys. Fluids {\bf 9}, 2644 (1997).

\bibitem{tsubota} 
M. Tsubota, K. Fujimoto, S.Y. Tsubota, J. Low. Temp. Phys. {\bf 188}, 119 (2017).

\bibitem{escartin} 
J. M. Escartin, F. Ancilotto, M. Barranco, and M. Pi Phys. Rev. B {\bf 99}, 140505(R) (2019).

\bibitem{mulker}
B.C. Mulkerin, R.M.W. van Bijnen, D.H.J. Or'Dell, A.M. Martin and N.G. Parker,
Phys. Rev. Lett. {\bf 111}, 170402 (2013).

\end{thebibliography}
\end{document}